\documentclass[sigconf]{acmart}

\usepackage[utf8]{inputenc} 
\usepackage[T1]{fontenc}    
\usepackage{hyperref}       
\usepackage{url}            
\usepackage{booktabs}       
\usepackage{amsfonts}       
\usepackage{nicefrac}       
\usepackage{microtype}      
\usepackage{lipsum}
\usepackage{graphicx}
\usepackage{amsmath}
\usepackage{color}
\usepackage{multirow}
\usepackage{bm}
\usepackage{amsmath}
\usepackage[lined, ruled, commentsnumbered, linesnumbered]{algorithm2e}
\usepackage{subcaption}
\SetKwRepeat{Do}{do}{while}

\definecolor{gray}{RGB}{192,192,192}
\definecolor{MyPink}{RGB}{255,178,178}
\definecolor{MyBlue}{RGB}{178,178,255}

\copyrightyear{2022}
\acmYear{2022}
\setcopyright{acmlicensed}\acmConference[SIGIR '22]{Proceedings of the 45th International ACM SIGIR Conference on Research and Development in Information Retrieval}{July 11--15, 2022}{Madrid, Spain}
\acmBooktitle{Proceedings of the 45th International ACM SIGIR Conference on Research and Development in Information Retrieval (SIGIR '22), July 11--15, 2022, Madrid, Spain}
\acmPrice{15.00}
\acmDOI{10.1145/3477495.3531941}
\acmISBN{978-1-4503-8732-3/22/07}

\begin{document}
\fancyhead{}

\title{AutoLossGen: Automatic Loss Function Generation for Recommender Systems}

\settopmatter{authorsperrow=4}

\author{Zelong Li}

\affiliation{
  \institution{Rutgers University}
  \city{New Brunswick, NJ, US}
}
\email{zelong.li@rutgers.edu}

\author{Jianchao Ji}
\affiliation{
  \institution{Rutgers University}
  \city{New Brunswick, NJ, US}
}
\email{jianchao.ji@rutgers.edu}

\author{Yingqiang Ge}
\affiliation{
  \institution{Rutgers University}
  \city{New Brunswick, NJ, US}
}
\email{yingqiang.ge@rutgers.edu}

\author{Yongfeng Zhang}
\affiliation{
  \institution{Rutgers University}
  \city{New Brunswick, NJ, US}
}
\email{yongfeng.zhang@rutgers.edu}

\def\authors{Zelong Li, Jianchao Ji, Yingqiang Ge, Yongfeng Zhang}

\begin{abstract}
In recommendation systems, the choice of loss function is critical since a good loss may significantly improve the model performance. However, manually designing a good loss is a big challenge due to the complexity of the problem.
A large fraction of previous work focuses on handcrafted loss functions, which needs significant expertise and human effort. In this paper, inspired by the recent development of automated machine learning, we propose an automatic loss function generation framework, AutoLossGen, which is able to generate loss functions directly constructed from basic mathematical operators without prior knowledge on loss structure.
More specifically, we develop a controller model driven by reinforcement learning to generate loss functions, and develop iterative and alternating optimization schedule to update the parameters of both the controller model and the recommender model. One challenge for automatic loss generation in recommender systems is the extreme sparsity of recommendation datasets, which leads to the sparse reward problem for loss generation and search. To solve the problem, we further develop a reward filtering mechanism for efficient and effective loss generation.
Experimental results show that our framework manages to create tailored loss functions for different recommendation models and datasets, and the generated loss gives better recommendation performance than commonly used baseline losses. Besides, most of the generated losses are transferable, i.e., the loss generated based on one model and dataset also works well for another model or dataset. Source code of the work is available at \url{https://github.com/rutgerswiselab/AutoLossGen}.
\end{abstract}



\maketitle

\section{Introduction}
\label{sec:introduction}
In this era of information explosion, recommendation system (RS) has become an important platform to filter unrelated items and to provide users with items of personalized interest.
Many researches are dedicated to optimizing RS models in order to promote the recommendation accuracy. 
However, a complete RS architecture consists of two vital parts: the RS model, and the loss function to optimize the RS model. Compared to the vast amount of research efforts on developing various kinds of RS models, the research on learning good loss function is still in its initial stage.

Actually, the choice of loss function may significantly influence the accuracy of the recommendation model. This is because the training of a RS model eventually depends on minimizing the loss function, and the gradient of the loss function supervises the optimization direction of the RS model. As a result, any inconsistency between the optimization goal and the optimization direction may hurt the model performance. An intuitive solution to this problem is directly using the optimization goal (e.g., the final evaluation metric) as the loss function. This can be effective when the evaluation metric is differentiable such as the root mean square error (RMSE) for some regression tasks. However, many metrics are non-differentiable and it is 
difficult to find derivatives such as the Area under the Curve (AUC) for classification tasks. For these cases, it is a challenge to design a good surrogate loss function to approximate the optimization goal, which needs comprehensive analysis and understanding of the task. Thanks to the meticulous design of researchers with their expertise and efforts, we have many useful handcrafted loss functions to solve the optimization problems under non-differentiable metrics \cite{rubinstein1999cross, hariharan2010large, lee2013study, liu2016large, lin2017focal, liu2017sphereface, masnadi2008design, masnadi2010design}.

Even though handcrafted loss functions have been used in various scenarios, it is still considerably beneficial to have methods that can automatically search and generate good loss functions. This is mainly for two reasons: 1) automatic loss generation helps to remove or reduce the manual efforts in loss design, and 2) the best loss could be different for different RS models and datasets, as a result, automatic loss generation can help to generate the best loss tailored to a specific model-dataset combination. 

In recent years, we have witnessed the development of automated machine learning (AutoML) techniques and especially neural architecture search (NAS) \cite{elsken2017simple, huang2018gnas, real2019regularized}, which can automatically design model architectures that are on par with or surpass the manually designed model architectures.
Inspired by the success of AutoML, we propose an automatic loss function generation (AutoLossGen) framework that can automatically generate loss functions for model optimization. AutoLossGen is different from existing loss learning research \cite{xu2018autoloss, li2019lfs, wang2020loss, liu2020stochastic, li2020auto, liu2021loss, li2021autoloss, zhao2021autoloss} on two perspectives. First, AutoLossGen is particularly designed for recommender systems, which present unique challenges due to the extreme data sparsity of recommender systems. This leads to the sparse reward problem in automatic loss generation, and to solve the problem, we propose a reward filtering mechanism for efficient and effective loss generation. Second, previous work on loss learning for recommender systems mostly focuses on automatic loss combination \cite{zhao2021autoloss}, which adopts several handcrafted base losses and learns the weight/importance for each loss, and then combines the losses through weighted sum as the final loss function. Different from previous work, we do not assume any prior knowledge on the loss structure, instead, we directly assemble new loss functions based on very basic mathematical operators, which can help to generate completely new loss functions. Our work is complementary to rather than adversary with previous loss combination methods because our generated new losses can be used as base losses, which can be combined with other handcrafted losses for better loss combination.

This paper makes the following key contributions:
\begin{itemize}
    \item We propose the AutoLossGen framework, which can generate loss functions directly constructed from basic mathematical operators without prior knowledge on loss structure.
    \item We develop proxy test and reward filtering mechanisms to speed up the generation process and mitigate the issues caused by the sparsity of RS datasets, so that the framework can produce reliable outcomes efficiently.
    \item We conduct experiments on two real-world datasets to show that the loss functions generated from the AutoLossGen framework outperform handcrafted base losses.
    \item We verify the transferability of our generated loss functions by showing that when applying them to other model-dataset settings, the losses can still achieve satisfactory performance.
\end{itemize}

In the following part of this paper, we first introduce the related work in Section \ref{sec:related_work}. In Section \ref{sec:framework}, we show the high-level architecture of our AutoLossGen framework. In Section \ref{sec:process}, we introduce the detailed design of the loss generation process along with methods for better generation efficiency. We provide and analyze the experimental results in Section \ref{sec:experiment}, and finally conclude the work together with future directions in Section \ref{sec:conclusions}.

\section{Related Work}
\label{sec:related_work}
In this section, we first introduce related work on automated machine learning (AutoML), and then we introduce loss learning which is a sub-category of AutoML.

\subsection{Automated Machine Learning}
Automated Machine Learning (AutoML) has been an important direction in recent years, which aims for reducing or even removing the requirement of human intervention in machine learning tasks.

There are three typical applications of AutoML \cite{yao2018taking}: 1) automated model selection, such as Auto-sklearn \cite{NIPS2015_11d0e628} and Auto-WEKA \cite{kotthoff2019auto}, which automatically selects a good machine learning model based on a library of models and hyper-parameter setting; 2) automated feature engineering, such as Data Science Machine \cite{kanter2015deep}, ExploreKit \cite{katz2016explorekit} and VEST \cite{cerqueira2021vest}, which generates or selects some useful features without manual intervention. Feature engineering is of great importance due to its great influence on model performance \cite{mitchell1997machine}; 3) neural architecture search (NAS), such as ENAS \cite{pham2018efficient}, DARTS \cite{liu2018darts}, NASH \cite{elsken2017simple}, GNAS \cite{huang2018gnas} and AmoebaNet-A \cite{real2019regularized}, which enables to search an effective neural network for a given task without manual architecture design. Experiments have shown that networks generated from NAS are on par with or surpass human-crafted architectures in different tasks.

Our work is related to NAS among these three applications. A loss for model training is usually a function that can be described as terms and operators, while searched architectures from NAS can be described as computation cells and connections. Thus, we can leverage the idea of NAS to construct a loss function search model implemented by reinforcement learning (RL). Although some research employs evolutionary algorithm \cite{real2017large, real2019regularized} and hill-climbing procedure \cite{elsken2017simple} for neural architecture search, RL has been shown effective in more research works, including \cite{zoph2018learning, liu2018progressive, baker2016designing, bello2017neural, cai2018path, pham2018efficient}, and thus becomes the dominant method in this field. To the best of our knowledge, we are the first to develop automated loss function generation frameworks by RL for recommender systems (RS).

\vspace{-2ex}
\subsection{Loss Learning}

Loss function plays an important role in machine learning, as it provides the direction for model training and significantly affects the performance \cite{rosasco2004loss}. Thus, besides model design, the choice of loss functions attracts more and more attention for specific tasks. Before the use of AutoML, loss functions are highly handcrafted and those handcrafted losses are shown to be effective and transferable under different scenarios. For regression tasks, mean absolute error (MAE) and root mean square error (RMSE) \cite{allen1971mean} are often employed in model evaluation \cite{willmott2005advantages, chai2014root}, and besides L1 and L2 losses, there are some loss variants including Smooth-L1 loss \cite{girshick2015fast}, Huber loss \cite{huber1992robust} and Charbonnier loss \cite{charbonnier1994two} for corresponding metrics. For classification tasks, since some metrics are non-differentiable, e.g., the Area under the ROC Curve (AUC) \cite{calders2007efficient}, more attempts on loss functions are made, including cross entropy (CE) \cite{rubinstein1999cross}, hinge loss \cite{hariharan2010large} and its variants \cite{lee2013study}, softmax loss and its variants \cite{liu2016large, liu2017sphereface}, Focal loss \cite{lin2017focal}, Savage loss \cite{masnadi2008design} and tangent loss \cite{masnadi2010design}.

With recent development of AutoML, some researchers propose and study automated loss learning to avoid the significant requirement of human efforts and expertise in loss design.
Xu et al. \cite{xu2018autoloss} design a framework to automatically select which loss to use and what parameters to update at each stage of the iterative and alternating optimization schedule. Li et al. \cite{li2019lfs} and Wang et al. \cite{wang2020loss} investigate the softmax loss to create an appropriate search space for loss learning and apply RL for the best parameter of the loss function. Liu et al. \cite{liu2020stochastic} provide a framework to automatically learn different weights of loss candidates for given data examples in a differentiable way. Li et al. \cite{li2020auto} substitute non-differentiable operators in metrics with parameterized functions to automatically search surrogate losses. Although these methods aim to learn losses automatically, they still depend on human expertise in the loss search process to a large extent, because the search process starts from existing loss functions. Liu et al. \cite{liu2021loss} and Li et al. \cite{li2021autoloss} search loss functions composed of primitive mathematical operators for several computer vision tasks by evolutionary algorithm, which is similar to our work, but we focus on different fields (recommender system) by the RL method and aim to address distinct challenges since the sparsity of recommender system datasets causes sparse reward issues if RL is directly applied. In the field of loss learning for RS, Zhao et al. \cite{zhao2021autoloss} propose a framework to search for an appropriate loss for a given data example, which adopts a set of base loss functions and dynamically adjust the weight of these loss functions for loss combination. Our method is different from and complementary to their work since we focus on generating new losses instead of combining existing losses. More specifically, we construct a new loss function starting from basic variables and operators without prior knowledge of loss structure or predefined loss functions. 

\section{Overall AutoLossGen Framework}
\label{sec:framework}
In this section, we show the high-level architecture and the main components of the AutoLossGen framework. We will introduce the refined details of the key components in the next section.

\begin{table}[t]
    \centering
    \begin{tabular}{c|l}
    \toprule
      {\bfseries Symbol} &{\bfseries Description}\\ 
      \midrule
      $U$ & The set of users in a recommender system\\
      $I$ & The set of items in a recommender system\\
      $u$ & A user ID in a recommender system\\
      $i$ & An item ID in a recommender system\\
      $\bm e_u$ & Embedding vector of the user $u$\\
      $\bm e_i$ & Embedding vector of the item $i$ \\
      $y_{ui}$ or $y$ & Ground-truth value of the pair $(u, i)$\\
      $\hat{y}_{ui}$ or $\hat{y}$ & Predicted value of the pair $(u, i)$\\
      $S$ & Set of controller's currently maintained variables\\
      $L$ or $L'$ & A list storing the generated candidate losses \\
      $B$ & A small batch of training data \\
      $\theta$ & Parameters of the controller model \\
      $\omega$ & Parameters of the recommender model \\
      $\mathcal{L}$ & Loss value \\
      $f$ & A sampled loss function from the controller model \\
      $\pi(f, \theta)$ & Policy of the controller model \\
      $\rho$ & Learning rate of the recommender model \\
      \bottomrule
    \end{tabular}
    \caption{Summary of the notations in this work.}
    \label{Table:notation}
    \vspace{-30pt}
\end{table}

\subsection{Problem Formalization and Notations}
\label{sec:problem}

Table \ref{Table:notation} introduces the basic notations that will be used in this paper.
In order to show our AutoLossGen framework is able to generate effective loss functions, we need to work on a concrete recommendation task so that evaluating the generated loss is possible.
In the following part of the paper, we explore the binary like/dislike prediction task for each user-item pair $(u,i)$ in recommender system. This can be formulated as either a 0-1 regression problem or a 0-1 classification problem, and we will explore both of them in the following.
To formalize the task, given a pair of user $u \in \bm U$ and item $i \in \bm I$, the learned RS model is required to accurately predict the likeness of user $u$ on item $i$ as $\hat{y}_{ui}$, while the ground-truth likeness is either $y_{ui}=1$ (like) or $y_{ui}=0$ (dislike). 
Following standard treatment on this problem, we consider the 5-star rating scale in this paper, while ratings $>3$ are considered as likes ($y_{ui}=1$) and ratings $\le3$ are considered as dislikes ($y_{ui}=0$).

Our framework contains two components, an RS model and a loss generation model (also called as the controller). These two parts are introduced in the following subsections.

\subsection{The Recommender System Model}
\label{sec:recsys_model}
\begin{figure}[t]
  \centering
  \includegraphics[width=0.47\textwidth]{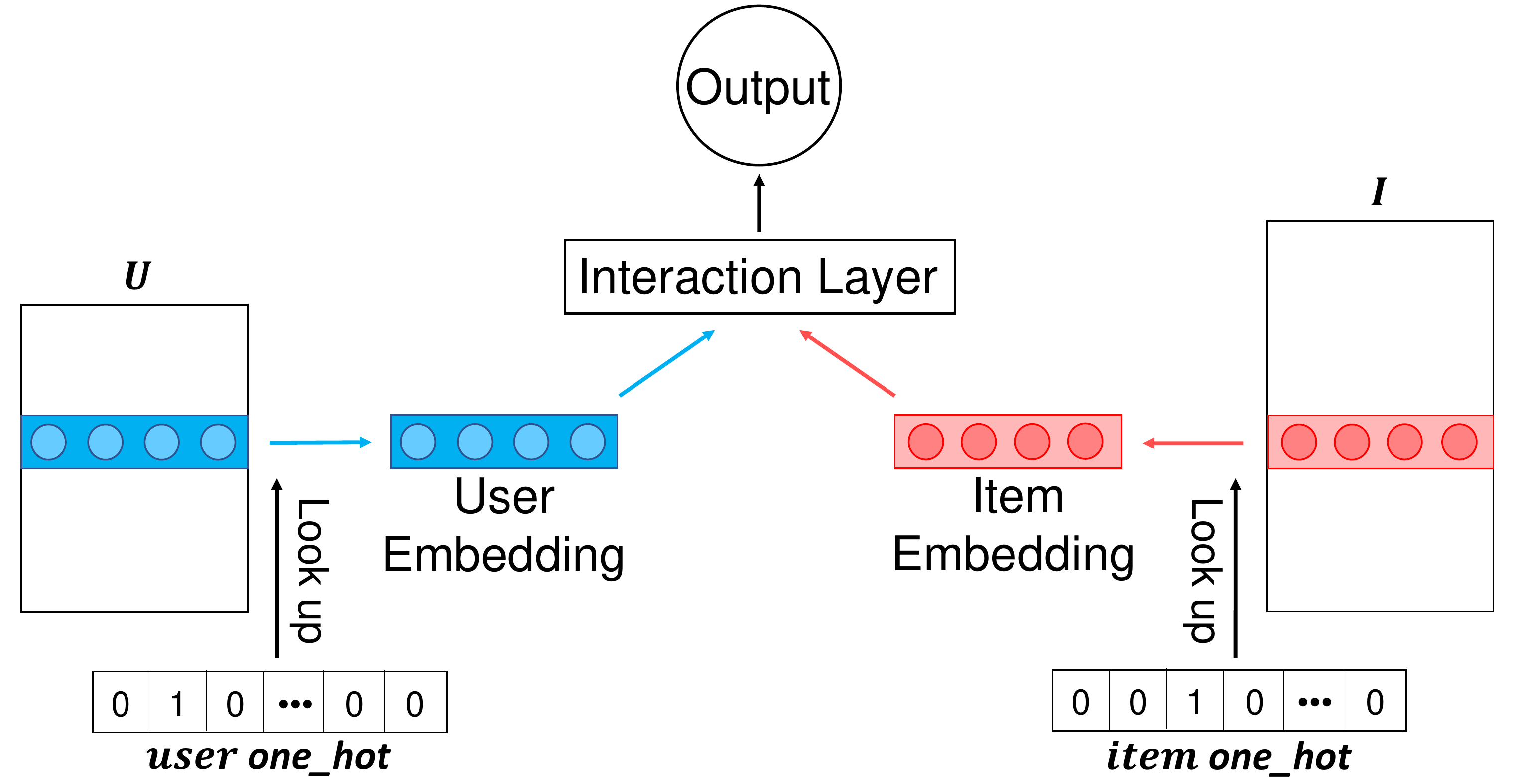}
  \vspace{-10pt}
  \caption{Structure of the recommender model}
  \vspace{-10pt}
  \label{fig:RSmodel}
\end{figure}

AutoLossGen framework is quite flexible for different kinds of recommender models. In this paper, we choose two simple but representative models as examples to explore, which are Matrix Factorization (MF) \cite{koren2009matrix} and the Multi-Layer Perceptron (MLP) network \cite{cheng2016wide} for recommendation. The former stands for traditional shallow matching method for recommendation, while the latter represents the deep matching model for recommendation. As shown in Figure \ref{fig:RSmodel}, the recommender model is composed of three layers: embedding layer, interaction layer and output layer, which are briefly introduced in the following.

\subsubsection{\textbf{Embedding Layer}}

In our recommendation task, we obtain the embedding vectors of users and items by their one-hot ID vectors. To formulate this, after transforming the user and item IDs to their corresponding one-hot vectors $u$ and $i$, the user embedding $e_u$ and item embedding $e_i$ are calculated as:
\begin{equation}
\begin{split}
    e_u = M_u \cdot u,~~~\text{and}~~~e_i = M_i \cdot i
\end{split}
\label{Eq:embedding}
\end{equation}
where $M_u \in \mathbb{R}^{d\times|U|}$ and $M_i \in \mathbb{R}^{d\times|I|}$ are the matrices storing the embeddings of all users $U$ and all items $I$, respectively, and they are learned during the training process. In this way, the sparse and high-dimensional one-hot vectors are compressed into low-dimensional embedding vectors, and we can retrieve their representations in this layer.

\subsubsection{\textbf{Interaction Layer}}

After the embedding layer, we feed the user and item representations into the interaction layer to make predictions. The structure of this layer is the main difference between MF and MLP. In MF model, we calculate the inner product of vectors with the bias terms, as shown in Eq.\eqref{Eq:MF}, where $b_u$, $b_i$ and $b_g$ are the user bias term, item bias term and global bias term, respectively. Together with $e_u$ and $e_i$, they are the parameters of MF to be learned.
\begin{equation}
    h_{out} = e_u \cdot e_i + b_u + b_i + b_g
    \label{Eq:MF}
\end{equation}

For MLP model, the structure of the interaction layer is multiple hidden layers composed of fully-connected layer and activation layer. The output of the $l$-th hidden layer is formulated as:
\begin{equation}
    h_{l+1} = ReLU(W_lh_l+b_l)
    \label{Eq:MLP}
\end{equation}
where $W_l$ is the weight matrix and $b_l$ is the bias vector, and the model uses $ReLU$ as the activation function. The input to the interaction layer of MLP is the concatenation of user and item embeddings, denoted as $h_0 = [e_u, e_i]$; the output is the result of the last activation layer, denoted as $h_{out} = h_n$ if there are $n$ hidden layers in total. Besides, to unify the format of the interaction layer, the dimension of the last hidden output layer is 1, i.e., $h_{out}$ is a real number. More implementation details are provided in the experiments.

\subsubsection{\textbf{Output Layer}}
The task of the output layer is to generate the final prediction $\hat{y}_{ui}$ of the RS model. Therefore, it could be different based on the range of $y$, but we can define the unified formula as:
\begin{equation}
    \hat{y}_{ui} = \sigma(h_{out})
    \label{Eq:output}
\end{equation}
where $\sigma(\cdot)$ is the activation function. For our task, it is \textit{sigmoid} since we limit the final output to a value between 0 and 1. 

\begin{table}[t]
    \centering
    \begin{tabular}{llc}
    \toprule
      {\bfseries Operator} &{\bfseries Expression} & {\bfseries Arity}\\ 
      \midrule
      Add & $x+y$ & 2 \\
      Multi & $x\cdot y$ & 2\\
      Max & max$(x, y)$ & 2\\
      Min & min$(x, y)$ & 2\\
      \midrule
      Neg & $-x$ & 1\\
      Identical & $x$ & 1\\
      Log & sign$(x)\cdot \text{log}(|x| + \xi)$ & 1\\
      Square & $x^2$ & 1\\
      Reciprocal & sign$(x) / (|x| + \xi)$ & 1\\
      \bottomrule
    \end{tabular}
    \vspace{5pt}
    \caption{The set of basic mathematical operators. $x$ and $y$ are variables for operators. $\xi = 10^{-6}$ is a small value to avoid numerical error such as division by zero. Besides, the result of each operation is clamped in the interval $[\xi, 1/\xi]$ to avoid numerical explosion.}
    \vspace{-20pt}
    \label{Table:operator}
\end{table}

\begin{figure}[t]
  \centering
  \includegraphics[width=0.47\textwidth]{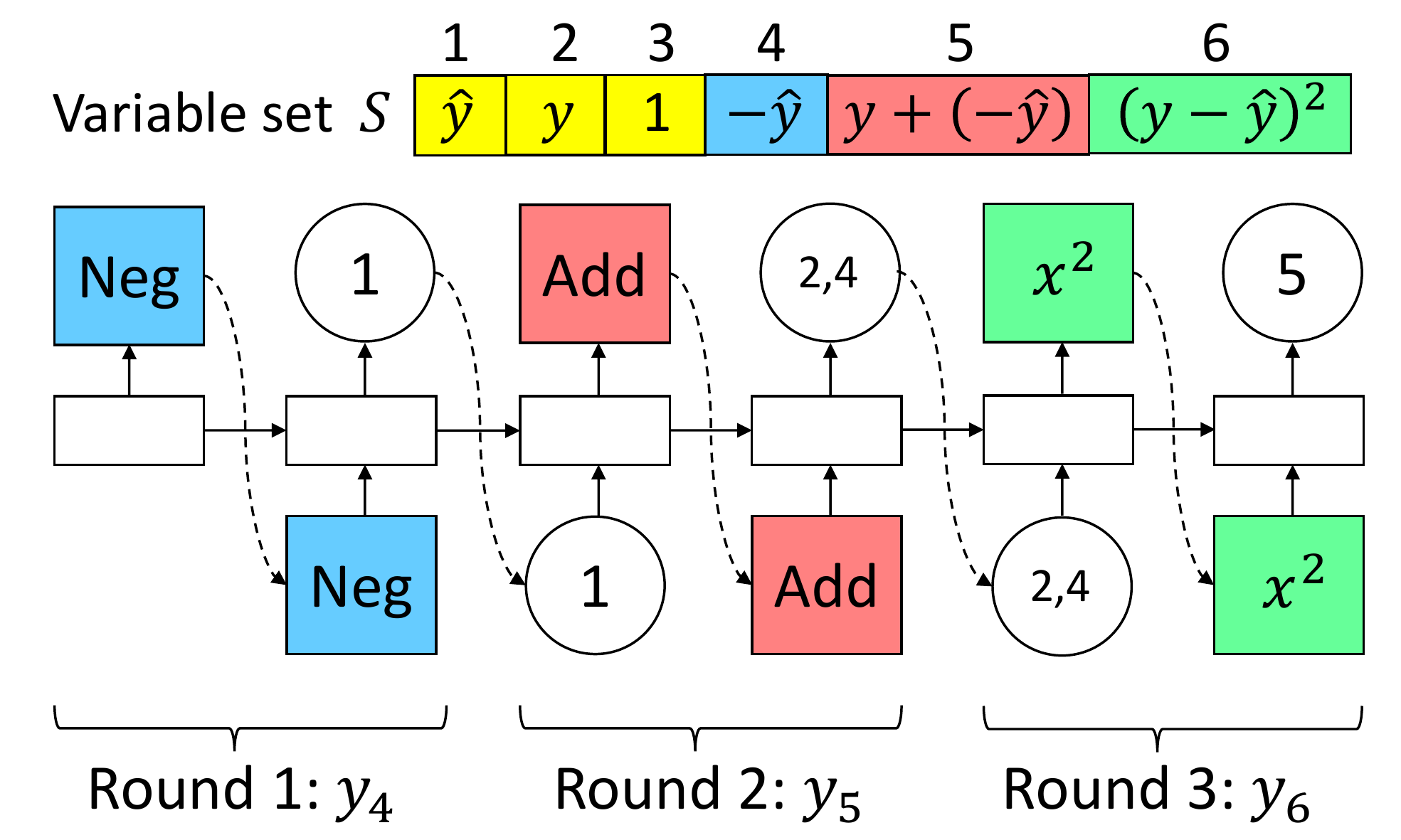}
  \caption{Taking the generation of MSE loss as an example to illustrate the loss generation process of the controller.}
  \vspace{-10pt}
  \label{fig:RNN}
\end{figure}

\subsection{The Controller Model}
\label{sec:controller}

The controller model is the most important part of the AutoLossGen framework, which is implemented as a recurrent neural network (RNN). The controller is able to generate various loss functions starting from basic mathematical operators. Our loss search space contains all possible functions composed of the operators shown in Table \ref{Table:operator}. Most common handcrafted losses are also built from these operators, such as the cross-entropy (CE) loss and L2 loss. Thus, we can expect that the search space is large enough for our controller to generate some effective loss functions.

The role of variables and operators in loss function is similar to the role of computation cells and edge connections in neural networks.
As a result, the loss generation process of RNN is similar to neural architecture search (NAS), shown as Figure \ref{fig:RNN}. At first, there are three initial variables in the variable set $S$: $\hat{y}$, $y$ and the number $1$. During each round, RNN samples an operator first, and then based on the arity of the operator shown in Table \ref{Table:operator}, samples the corresponding number of distinct input variables from the set $S$. For example, in Figure \ref{fig:RNN}, RNN samples \textit{Negative} operator in the first round, so for the next step, one variable from the variable set $S$ is sampled (suppose $\hat{y}$ is sampled), and a new variable (i.e., $-\hat{y}$) is created and added into the variable set $S$; in the second round, RNN samples \textit{Add} operator, and thus two different variables are needed to execute the operation. The distribution of operator and variable sampling in each round is determined by the RNN's hidden output vector from the previous round, which goes through a soft-max layer to create a sampling probability vector with the same size as the current number of candidate operators and candidate variables in $S$.
Besides, to encourage complex functions in fewer rounds, our framework allows variables to be used multiple times.

Finally, the controller takes the last variable in the variable set as the loss function.
One thing to note here is that there is no significant relationship between the complexity and the performance of loss functions, i.e., the most effective loss function could have either complex or concise mathematical forms.
As a result, we would not like to apply too many restrictions during the loss sampling process, and because of this, our framework is not only able to generate complex loss functions with various forms, but also includes short expressions in the search space. More precisely, the loss function search space includes every possible expression based on the pre-defined operators that utilizes at most $m$ intermediate variables, where $m$ is defined as the maximum number of rounds in RNN. However, the flexibility of sampled loss functions also brings new problems, including the zero-gradient problem and the duplicated function problem, and some bad functions may ruin the performance if used as loss. We will provide solutions to these problems in the loss generation process, specifically in Section \ref{sec:loss search}.

\vspace{-4pt}
\section{Loss generation process}
\label{sec:process}

\begin{figure}[t]
  \centering
  \includegraphics[width=0.47\textwidth]{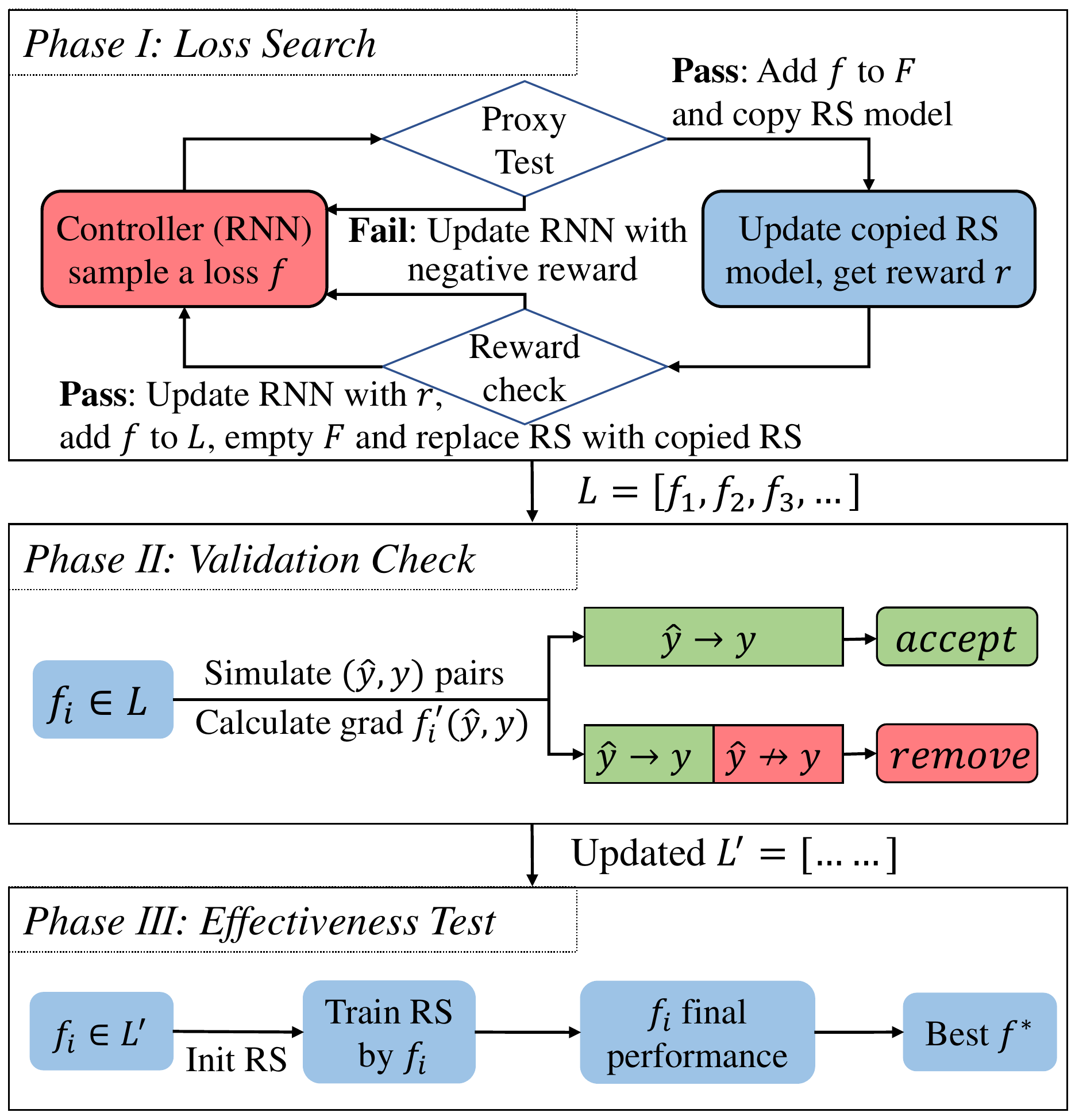}
  \vspace{-10pt}
  \caption{Overview of the loss generation process with three phases: loss search, validation check, and effectiveness test.}
  \vspace{-10pt}
  \label{fig:process}
\end{figure}

There are three phases when executing our AutoLossGen framework, as shown in Figure \ref{fig:process}. In Phase I (loss search), we optimize all parameters of our framework in iterative and alternating schedule, and record the loss function we generated in every RL optimization loop. 
In Phase II (validation check), we remove those functions that cannot produce correct gradient direction for simulated $(y,\hat{y})$ pairs.
In Phase III (effectiveness test), for each selected loss in Phase II, we randomly initialize a recommender model and train the model to convergence using the loss to obtain the final performance for the loss, and we keep the best performance loss as the finally selected loss function. In the following part of this section, we introduce the loss generation phases with the proposed techniques in detail.

\subsection{Loss Search Phase}
\label{sec:loss search}

\subsubsection{\textbf{Iterative and Alternating Optimization Schedule}}

As introduced in Section \ref{sec:framework}, both parameters in the recommender model $\omega$ and parameters in the RNN controller model $\theta$ need optimization in the search process. For the RS model, we perform stochastic gradient descent (SGD) on it to update $\omega$, while for the controller, inspired by the success of neural architecture search (NAS) in \cite{zoph2016neural, pham2018efficient}, we apply the REINFORCE \cite{williams1992simple} algorithm to update $\theta$, and the performance increment on the validation dataset is used as the reward signal for policy gradient. We use the Area under the ROC Curve (AUC) for classification task and root mean square error (RMSE) for regression task to evaluate the performance increment and obtain the reward on validation set since they are prediction-sensitive, i.e., the metric values will be different with a very small fluctuation on predictions \cite{allen1971mean, calders2007efficient}, so that non-trivial reward can be calculated for better update on the controller. 

The search process is a typical bi-level optimization problem \cite{anandalingam1992hierarchical}. Ideally, after the controller samples a new loss function, we cannot judge the performance of the loss and retrieve the reward signal until the RS model is updated to convergence. The optimization problem can be formulated as:
\begin{equation}
\begin{split}
    \max\limits_{\theta} \text{ } & \mathbb{E}_{f\sim\pi(f, \theta)}[\mathcal{R}_{val}(f, \omega^*(f))] \\
    \text{s.t.~~~} & \omega^*(f) = \text{arg}\min\limits_{\omega}\mathcal{L}_{train}(\omega, f)
\end{split}
\label{Eq:optimization}
\end{equation}
where $\mathbb{E}[\mathcal{R}_{val}]$ represents the expected reward on validation sets, and other symbols can be referred in Table \ref{Table:notation}. However, due to the large search space, the nested optimization process is too time-consuming to be put into practice. Referring to DARTS techniques \cite{liu2018darts}, we utilize first-order approximation of the gradient for the RS model, and leverage the temporary reward to update the controller. To be specific, we update the RS model with one epoch of training data by the sampled loss function to approximate the optimization effect, shown as Eq.\eqref{Eq:approximate}, and then use the performance increment of the current RS model on validation datasets as the reward of the sampled loss to update the controller. The process is called iterative and alternating optimization, since the updates of the RS model and the controller is alternating in each iteration. The pseudo code of our optimization method is described in Algorithm \ref{algorithm:update}. In Section \ref{sec:experiment}, we experimentally show that the first-order approximation is able to help our framework generate effective loss functions.
\begin{equation}
    \omega^*(f) \approx \omega - \rho\nabla_{\omega}\mathcal{L}_{train}(\omega, f)
\label{Eq:approximate}
\end{equation}

\setlength{\textfloatsep}{9pt}
\begin{algorithm}[t]
    \textbf{Input: } Controller, RS model, Reward threshold $\eta$ \\
    \textbf{Output: } $[f_1, f_2, f_3...]$ \\
    \While {not converge} {
        $init \leftarrow$ performance of RS on validation data \\
        \Do{$f$ fails on the proxy test}{
            Controller generates a loss $f(\hat{y}, y, 1)$ \\
        }
        Copy current RS model \\
        Update the copied RS by $f$ with one epoch of train data \\
        $updated$ $\leftarrow$ performance of copied RS on validation data \\
        $reward \leftarrow updated - init$ \\
        \If {$reward \ge -\eta$} {
            Replace RS model by copied RS \\
            Record $f$ into $L$ \\
        }
    }
    \Return $L$ \\
    \caption{\mbox{Iterative and alternating optimization schedule}}
    \label{algorithm:update}
\end{algorithm}

\subsubsection{\textbf{Proxy Test}}
\label{sec:proxy test}

In Algorithm \ref{algorithm:update}, we mentioned a proxy test after the controller samples a loss. As analyzed in Section \ref{sec:controller}, we expect fewer restrictions on the form of losses when sampling functions, and as a side effect, zero-gradient functions and gradient-level duplicated functions may appear during the search process since the form of sampled functions can be very simple. Zero-gradient functions are the functions whose gradients over $\hat{y}$ are always zero, such as $(y + 1)$, and gradient-level duplicated functions are the functions that have the same or very close gradient values over $\hat{y}$ with already sampled functions. For instant, if MSE loss $(\hat{y} - y)^2$ is already sampled from our framework in the current RL optimization loop, then $(\hat{y} - y)^2 + y$ will be considered as a duplicated loss function since its gradient over $\hat{y}$ is the same as MSE and thus their effects are equivalent in gradient-based optimization. To efficiently skip these losses, at the beginning of the whole loss search process, we sample a small batch of training data $B$, where $|B|$ is set between 5 to 20 according to the size of datasets and $B$ will not be changed. After the controller samples a loss $f$, we first calculate the gradient of the RS model on $B$, denoted as $\nabla_\omega(B, f)$, and if:
\begin{equation}
    \|\nabla_{\omega}(B, f)\| < \delta
\label{Eq:zero-gradient}
\end{equation}
where $\delta$ is a small value set as $10^{-4}$ in implementation, then we treat $f$ as a zero-gradient loss and provide a default negative reward to update the controller since we do not want zero-gradient losses in future rounds. Besides, we use a set $F$ to temporarily store the already sampled losses in the current RL optimization loop, and if:
\begin{equation}
    \exists f'\in F\text{ ,  s.t. }\|\nabla_{\omega}(B, f) - \nabla_{\omega}(B, f')\| < \delta
\label{Eq:duplicate}
\end{equation}
then we consider $f$ as a duplicated loss and directly use the reward of $f'$ to update the controller. 
If $f$ is not considered as zero-gradient or duplicated losses (i.e, if $f$ passes the proxy test in Eq.\eqref{Eq:zero-gradient} and \eqref{Eq:duplicate}), 
then we add $f$ into $F$ and calculate its reward over RS model. 
In this way, zero-gradient and gradient-level duplicated losses are quickly skipped after proxy test and thus we do not have to waste time training the RS model over such losses.

\subsubsection{\textbf{Reward Filtering Mechanism}}
\label{sec:reward filtering}
During the training process of the RS model, in most cases, we would have to use more steps to correct the model if the model is trained along the wrong direction by a bad loss. As a result, to speed up the loss generation process and avoid degradation on RS model performance, we do not directly test a generated loss on the RS model, instead, we make a copy of the RS model and optimize the copied model over a generated loss to calculate the reward of the loss. Besides, we only replace the RS model with the copied RS model if $reward \ge -\eta$, otherwise, we discard the copied RS model and do not update the parameter of the RS model.
Here, we allow some minor negative rewards to 
provide RL with some exploration ability on top of exploitation so as to avoid getting stuck in local optima. In Section \ref{sec:ablation}, we will show through ablation study that without the reward filter mechanism, the performance of RS model may be unstable and no effective loss function can be generated during the search.

\subsection{\mbox{Validation Check and Effectiveness Test}}

\begin{figure}[t]
\begin{minipage}{0.23\textwidth}
  \centering
  \includegraphics[width=\linewidth]{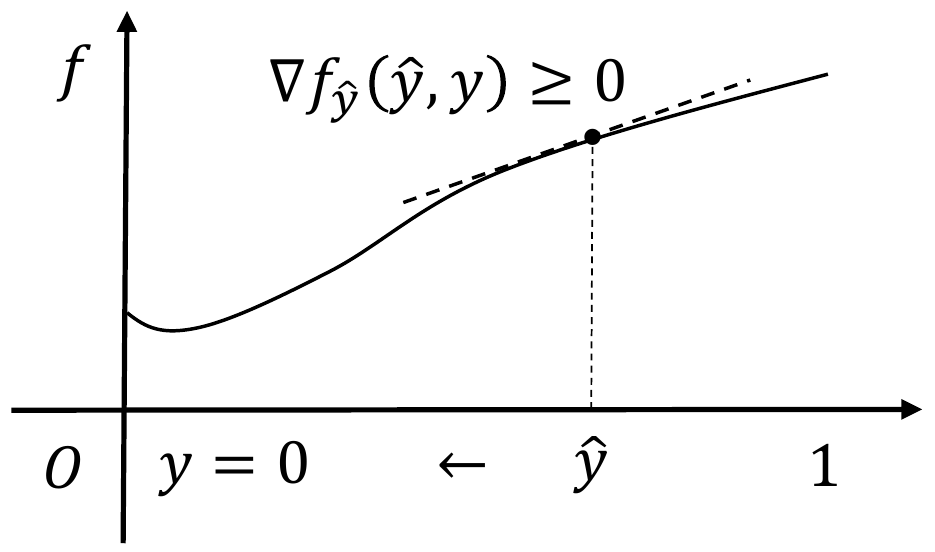}
  \vspace{-10pt}
  \subcaption{$y=0$}
  \label{fig:valid check y0}
\end{minipage}\hfill
\begin{minipage}{0.23\textwidth}
  \centering
  \includegraphics[width=\linewidth]{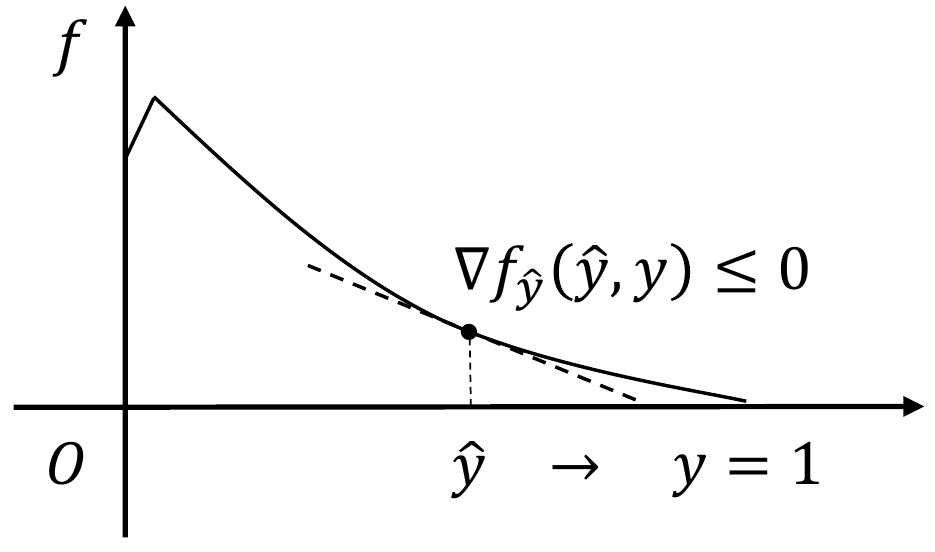}
  \vspace{-10pt}
  \subcaption{$y=1$}
  \label{fig:valid check y1}
\end{minipage}
\vspace{-10pt}
\caption{A toy example to show how a loss function can pass the validation check.}
\label{fig:valid check}
\end{figure}

We expect our generated losses can update a randomly initialized RS model from the beginning to convergence throughout the whole training process. However, the output loss functions from the loss search phase are only tested to be effective for a certain epoch of optimizing the RS model. Loss functions should perform well for various $y$ and $\hat{y}$ values in the domain of definition of the values.
Out of such consideration, we design a validation check phase to filter the loss functions provided by the Loss Search Phase.

More specifically, we sample different values of $\hat{y}$ and $y$ from $\hat{y}\in [0, 1]$ and $y\in \{0, 1\}$ and create a set of synthesized $(\hat{y}, y)$ pairs. For each candidate loss function $f$ from the previous phase, we calculate the gradient of $f$ over $\hat{y}$, denoted as $\nabla f_{\hat{y}}(\hat{y}, y)$. If $\nabla f_{\hat{y}}(\hat{y}, y) \ge 0$ when $y=0$ and $\nabla f_{\hat{y}}(\hat{y}, y) \le 0$ when $y=1$ for all of the synthesized pairs, then the loss function is considered valid, otherwise, it is removed from the loss candidate set. As shown in Figure \ref{fig:valid check}, the intuition is that if the ground truth label is $y=0$, then we hope the gradient direction on $\hat{y}$ is positive, and because we use the inverse gradient direction for loss minimization, so $\hat{y}$ will be optimized towards 0 during optimization. Similar for the case of $y=1$. Based on this, we can filter out those losses whose gradients are infeasible for optimization.

Finally, in Phase III, we leverage each candidate loss to train a randomly initialized RS model to convergence and use the model performance on validation set as the effectiveness of the candidate loss, and we take the most effective loss as the final output loss of the whole loss learning process. In the experiment part, we will report the performance of the generated loss on test set.

\section{Experiments}
\label{sec:experiment}

In this section, we conduct experiments to evaluate the effectiveness of generated loss functions
and to help better understand the loss generation process.\footnote{Source code available at \url{https://github.com/rutgerswiselab/AutoLossGen}.}

\subsection{Experimental Setup}

\subsubsection{\textbf{Dataset Description}}

\begin{table}[t]
    \centering
    \begin{tabular}{c|rrrrr}
        \toprule
         Dataset & \#Users & \#Items  & \#Pos & \#Neg & Density\\
         \midrule
         ML-100K & 943 & 1,682 & 55,375 & 44,625 & 6.30\%\\
         Electronics & 192,403 & 63,001 & 1,356,067 & 333,121 & 0.014\%\\
         \bottomrule
    \end{tabular}
    \vspace{5pt}
    \caption{Basic statistics of the datasets}
    \vspace{-10pt}
    \label{Table:dataset}
\end{table}

Our experiments are conducted on two widely-used benchmark datasets of recommender systems, namely, ML-100K and Amazon Electronics \cite{harper2015movielens,he2016ups,geng2022recommendation,ge2019maximizing,ge2020learning,fu2021hoops,ge2022explainable,xu2021causal}, which are both publicly available. The detailed statistics of the datasets are shown in Table \ref{Table:dataset}, and we briefly introduce these two datasets in the following part.

\begin{itemize}
\item \textbf{ML-100K}: It is a widely used RS dataset maintained by GroupLens.\footnote{\url{https://grouplens.org/datasets/movielens/100k/}} It contains 100,000 ratings from 943 users on 1,682 movies. The rating values are integers ranging from 1 to 5 (both included).

\item \textbf{Amazon Electronics}: This is one of the Amazon 5-core e-commerce datasets\footnote{\url{http://jmcauley.ucsd.edu/data/amazon/index.html}} that records the rating of items given by users on Amazon spanning from May 1996 to July 2014. We use the Electronics category with over one million interactions, which is larger and sparser than ML-100K. The rating values are integers ranging from 1 to 5 (both included).
\end{itemize}

\begin{table*}[t]
    \centering
    \setlength{\tabcolsep}{5pt}
    \begin{tabular}{c|c|c|c|c|c}
        \toprule
        Model & Dataset & Task & Loss Name & Abbr. & Loss Formula \\
        \midrule
        MF & ML-100K & \multirow{5.5}{*}{Classification} & Max of Ratio Loss & MaxR & max$\big((\hat{y} + \epsilon)/(y + \epsilon), (y + \epsilon)/(\hat{y} + \epsilon)\big)$ \\
        \cmidrule{1-2} \cmidrule{4-6}
        MF & Electronics & & \multirow{2}{*}{Sum of Reciprocal Loss} & \multirow{2}{*}{SumR} & \multirow{2}{*}{$(\hat{y} + y + \epsilon)/(\hat{y} \cdot y + \epsilon)$} \\
        \cmidrule{1-2}
        MLP & ML-100K & & & & \\
        \cmidrule{1-2} \cmidrule{4-6}
        MLP & Electronics & & Log Min Based Loss & LogMin & $\log\big((1 + \epsilon)/(\min(\hat{y}, y) + \epsilon)\big) \cdot \big(\hat{y} + y + \min(\hat{y}, y)\big)$ \\
        \midrule
        * & * & Regression & Mean Square Error Loss & MSE & $(\hat{y} - y)^2$ \\
        \bottomrule
    \end{tabular}
    \vspace{5pt}
    \caption{Loss generation result, $\epsilon$ is a tunable parameter representing the smoothing coefficient to prevent numerical errors such as division by zero. MSE loss is generated in all four model-dataset combinations for the regression task.}
    \vspace{-20pt}
    \label{Table:loss generation}
\end{table*}

As mentioned in Section \ref{sec:problem}, we would like to predict the explicit feedback from users, either as a classification task or as a regression task.
Following standard treatment, ratings $\ge4$ are considered as positive (like) with label as 1, while ratings $\le3$ are considered as negative (dislike) with label as 0. 
We use positive Leave-One-Out to create the train-validation-test datasets \cite{shi2020neural, chen2021neural,chen2022graph}. Specifically, for each user, based on timestamp, we put the user's last positive interaction, along with following negative interactions, into the test set, put the second-to-last positive interaction, together with remaining following negative interactions, into the validation set, and put all of the remaining interactions into the training set. If a user has fewer than 5 interactions, we put all its interactions into the training set to avoid the cold start problem.

\subsubsection{\textbf{Baseline Losses}}

We compare with the following baseline losses in the experiment.

\begin{itemize}
    \item \textbf{Mean Square Error (MSE)} \cite{allen1971mean}: MSE is a commonly used loss for regression tasks, but also shows good performance for classification. It minimizes the square of difference between label and prediction values.
    \item \textbf{Binary Cross Entropy (BCE)} \cite{rubinstein1999cross}: BCE is a special form of cross entropy (CE) for binary classification task. It is one of the most widely used loss functions for the task. Besides, two well-known losses, logistic loss and Kullback–Leibler (KL) divergence \cite{kullback1951information}, are different from BCE only in constants for binary classification task, and thus, we use BCE as a baseline loss to represent these types of losses.
    \item \textbf{Hinge Loss} \cite{hariharan2010large}: Hinge loss is a margin-based loss function. It does not require the prediction to be exactly the same as the true value. Instead, if $y$ and $\hat{y}$ are close enough, the loss value would be 0, which is reasonable for classification tasks.
    \item \textbf{Focal Loss} \cite{lin2017focal}: Focal is recently proposed and revised from the CE loss. It is designed for models to concentrate on hard samples and reduce the weight of well-classified samples.
\end{itemize}

We are aware of some recent loss combination techniques such as SLF \cite{liu2020stochastic} and AutoLoss \cite{zhao2021autoloss}. However, these models do not generate new loss functions. Instead, they learn a weighted sum of existing handcrafted losses as the final loss. However, our work aims to generate new individual losses rather than weighted combination of losses, as a result we only compare with individual loss functions. Actually, our work is complementary to rather than adversary with these loss combination methods, because our generated new losses can be used as base losses together with existing handcrafted losses for loss combination. As a result, our method will positively contribute to the loss combination methods if our generated loss functions are better than existing handcrafted loss functions.

\subsubsection{\textbf{Evaluation Metrics}}

To evaluate the final performance of the generated loss functions, we use the Area under the ROC Curve (AUC), F1-score and Accuracy for evaluating the classification task, and use mean absolute error (MAE) and root mean square error (RMSE) for evaluating the regression task.

\begin{table*}[t]
    \centering
    \setlength{\tabcolsep}{10.5pt}
    \begin{tabular}{c|c|c|cccc|ccc}
        \toprule
        \multirow{2.5}{*}{Model} & \multirow{2.5}{*}{Dataset} & \multirow{2.5}{*}{Metric} & \multicolumn{4}{c|}{Handcrafted Loss} & \multicolumn{3}{c}{Generated Loss} \\
        \cmidrule{4-10}
        & & & MSE & BCE & Hinge & Focal & MaxR
        & SumR 
        & LogMin
        \\
        \midrule \midrule
        \multirow{3}{*}{MF} & \multirow{3}{*}{ML-100K} & AUC $\uparrow$ & 0.7808 & 0.7882 & 0.7848 & 0.7930 & \bm{$0.8087^*$} & \bm{$0.8086$} & \bm{$0.7981$} \\ 
        & & F1-score $\uparrow$ & 0.6058 & 0.6073 & 0.6133 & 0.6121 & \bm{$0.6260^*$} & \bm{$0.6245$} & \bm{$0.6160$} \\
        & & Accuracy $\uparrow$ & 0.6919 & 0.6972 & 0.7239 & 0.7011 & \bm{$0.7305$} & \bm{$0.7645^*$} & \bm{$0.7398$} \\
        \midrule
        \multirow{3}{*}{MF} & \multirow{3}{*}{Electronics} & AUC $\uparrow$ & 0.6510 & 0.6515 & 0.6689 & 0.6521 & \bm{$0.6697^*$} & \bm{$0.6695$} & 0.6534 \\
        & & F1-score $\uparrow$ & 0.8843 & 0.8843 & 0.8846 & 0.8843 & 0.8846 & 0.8846 & 0.8844 \\
        & & Accuracy $\uparrow$ & 0.7927 & 0.7927 & 0.7937 & 0.7927 & 0.7937 & 0.7937 & 0.7927 \\
        \midrule \midrule
        \multirow{3}{*}{MLP} & \multirow{3}{*}{ML-100K} & AUC $\uparrow$ & 0.7655 & 0.7725 & 0.7472 & 0.7629 & \bm{$0.7747$} & \bm{$0.7743$} & \bm{$0.7752^*$} \\
        & & F1-score $\uparrow$ & 0.5938 & 0.5985 & 0.5865 & 0.5890 & \bm{$0.6130^*$} & \bm{$0.6053$} & \bm{$0.6064$} \\
        & & Accuracy $\uparrow$ & 0.6625 & 0.6797 & 0.6717 & 0.6437 & \bm{$0.7077^*$} & \bm{$0.6830$} & \bm{$0.6909$} \\
        \midrule
        \multirow{3}{*}{MLP} & \multirow{3}{*}{Electronics} & AUC $\uparrow$ & 0.6232 & 0.6228 & 0.6242 & 0.6205 & \bm{$0.6250$} & \bm{$0.6404^*$} & \bm{$0.6318$} \\
        & & F1-score $\uparrow$ & 0.8843 & 0.8843 & 0.8843 & 0.8843 & \bm{$0.8844^*$} & 0.8843 & 0.8843 \\
        & & Accuracy $\uparrow$ & 0.7926 & 0.7926 & 0.7926 & 0.7926 & \bm{$0.7928^*$} & \bm{$0.7927$} & 0.7926 \\
        \bottomrule
    \end{tabular}
        \vspace{5pt}
    \caption{Final performance on the classification task. $\uparrow$ means the measure is the higher the better. Bold numbers indicate its performance is significantly better at $\bm{p < 0.01}$ than all baseline losses, and * represents the best performance of each row.}
    \vspace{-20pt}
    \label{Table:result}
\end{table*}

\subsubsection{\textbf{Implementation Details}}

Our framework and all baselines are implemented by PyTorch, an open source library. 
As mentioned in Section \ref{sec:recsys_model}, we test on two types of RS models, a shallow matching model based on Matrix Factorization (MF) and a neural matching model based on Multi-Layer Perceptron (MLP). The implantation details of the RS models are as follow: (a) \textit{Embedding layer}: we set the dimension of embedding vectors as 64 for both users and items. (b) \textit{Interaction layer}: we do not have hyper-parameters for MF, and for the MLP model, we have three fully-connected layers with layer size $128\times64$, $64\times16$ and $16\times1$. For each layer of MLP, we leverage batch normalization, dropout (with rate as 0.2) and ReLU function for activation. (c) \textit{Output layer}: we use \textit{sigmoid} function since $y$ is either 0 or 1, and $\hat{y}$ is expected to be in-between 0 and 1.

Besides the RS model, another critical part of the AutoLossGen framework is the controller model. The controller RNN is implemented by a two-layer LSTM with 32 hidden units on each layer, and the weights of controller are uniformly initialized between -0.1 and 0.1. We use logit clipping with the tanh constant of 1.5 to limit the range of logits so as to control the sampling entropy. This can help to increase the sampling diversity and avoid premature convergence~\cite{bello2016neural}. We also add the controller's sampling entropy to the reward, weighted by 0.0001, to drive the sampling process towards a relatively stable status. The largest length of variable set is fixed to 10, i.e., our search space includes all functions that utilizes at most 10 intermediate variables including $y, \hat{y}$ and 1.

During the loss generation process, we use iterative and alternating schedule to optimize the controller and RS models. When training the RS model, we fix the parameters of the controller $\theta$, and use stochastic gradient descent (SGD) optimizer with learning rate as 0.01 to update the RS model; when training the controller, we fix the parameters of the RS model $\omega$, and use Adam optimizer with learning rate as 0.001. To reduce random bias, we average the rewards of ten sampled loss functions to update the controller. $\ell_2$ parameter regularization is adopted in both RS model and controller model optimization. For the proxy test in the loss search phase, the batch size $|B|$ is set as 5 for the smaller ML-100K dataset and 20 for the larger Electronics dataset.
We want the exploration process to be as thorough as possible, as a result, the termination condition we set is that the RS model converges or the controller cannot sample an effective loss to pass the reward filtering mechanism over 24 hours. The running time of the exploration process varies from about two days to one week with different model-dataset combinations. However, the exploration and loss learning process is like gold mining---once the loss function is found, we do not have to re-run the process any more.

For the validation check in the second phase of Figure \ref{fig:process}, the number of simulated pairs is $2,000$. In the effectiveness test phase, we randomly initialize a new RS model and use SGD for model optimization. There may exist tunable parameters to avoid numerical errors such as division by zero in the generated loss functions (e.g., the $\epsilon$ in Table \ref{Table:loss generation}), and we use grid search in [1, $10^{-1}$, $10^{-2}$, $10^{-3}$, $10^{-4}$, $10^{-5}$, $10^{-6}$] on validation set to decide the value of the parameters. To prevent over-fitting, if the performance of the RS model on validation set is decreasing in 10 consecutive epochs, or the best performance on validation set is over 50 epochs before, then the training process will be early terminated.

\begin{figure}[t]
\begin{minipage}{0.23\textwidth}
  \centering
  \includegraphics[width=\linewidth]{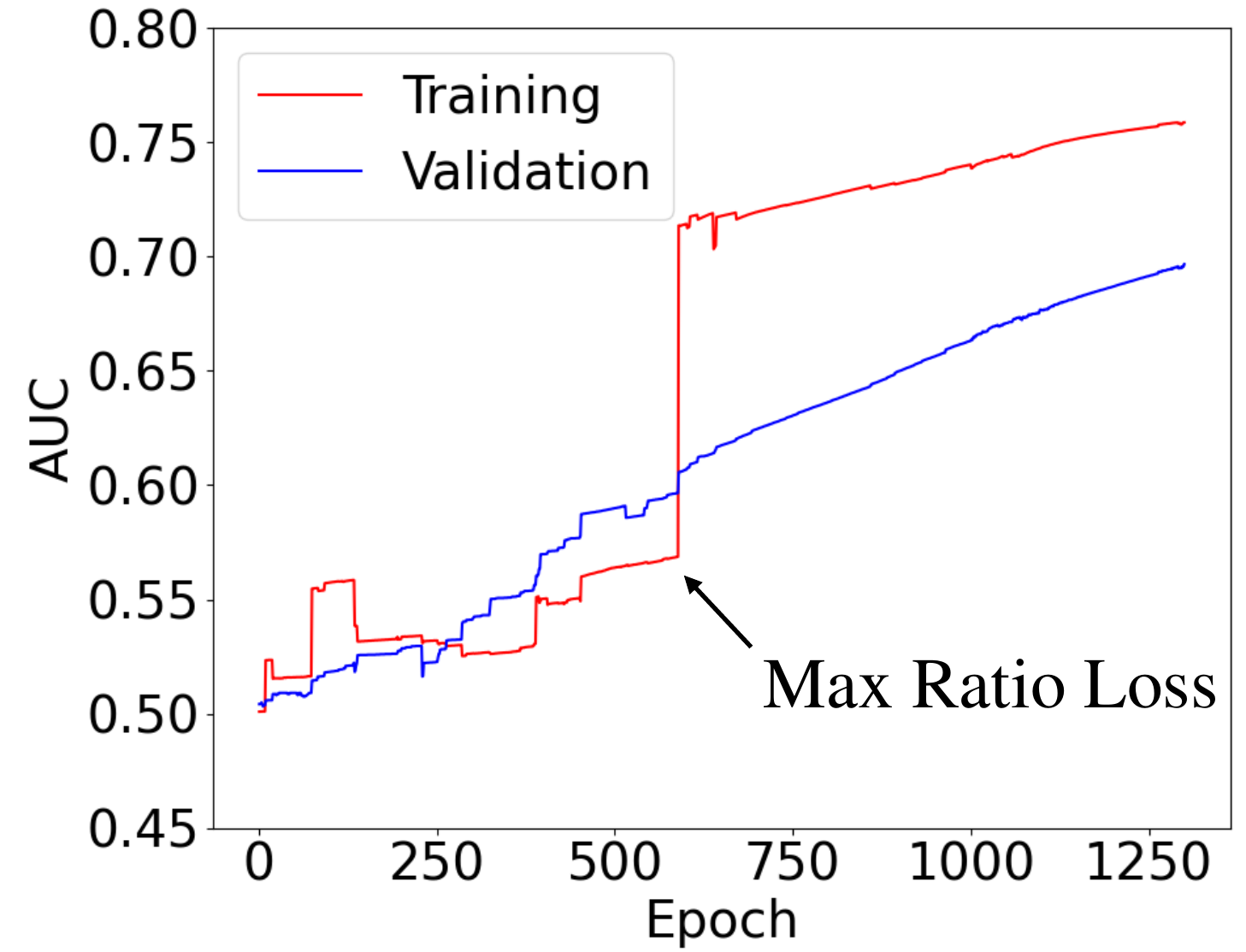}
  \subcaption{Complete AutoLossGen}
  \label{fig:process exp}
\end{minipage}\hfill
\begin{minipage}{0.23\textwidth}
  \centering
  \includegraphics[width=\linewidth]{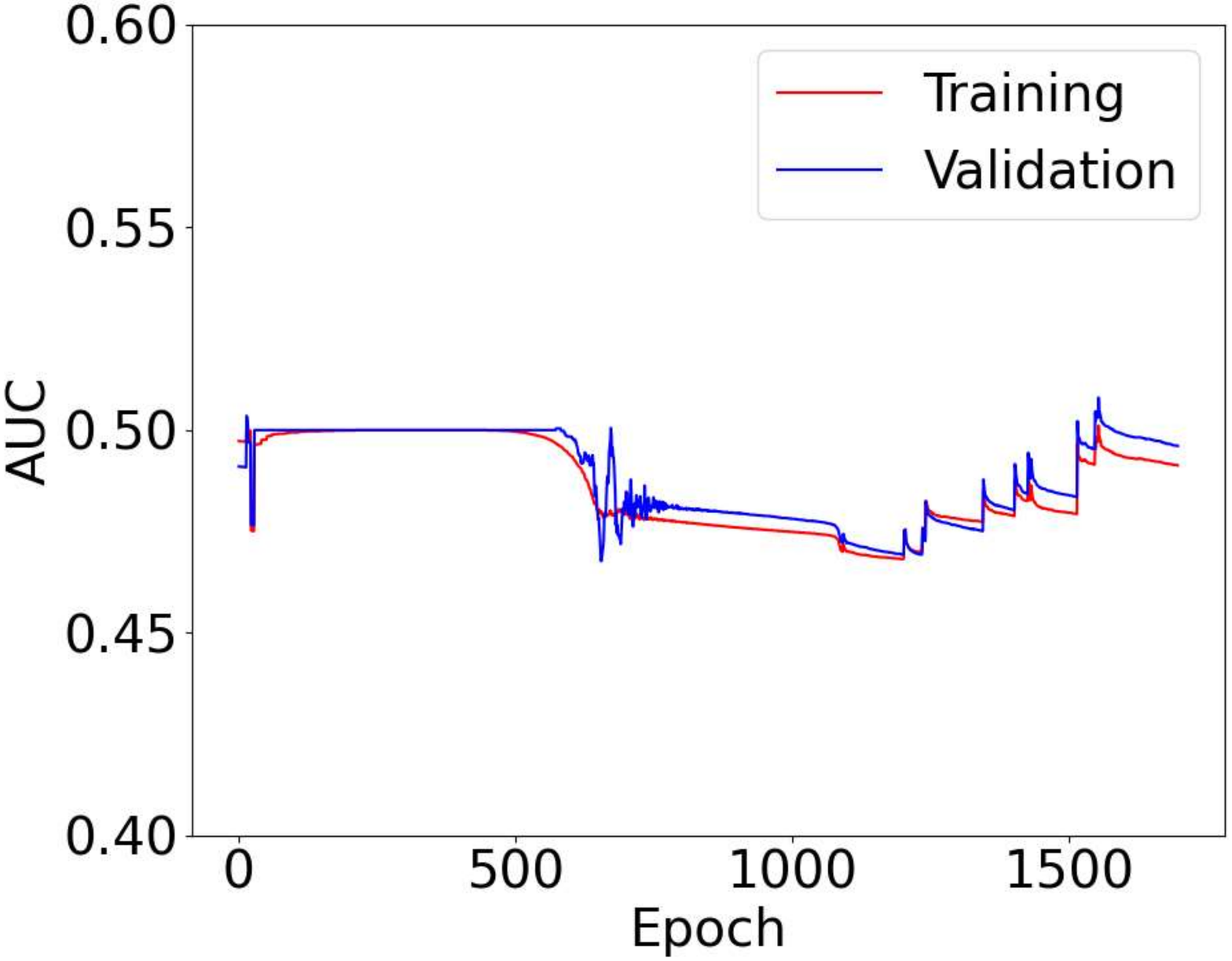}
  \subcaption{Without Reward Filtering}
  \label{fig:wandering}
\end{minipage}
\vspace{-10pt}
\caption{The performance of MF model on training and validation datasets of ML-100K during loss generation process under the classification task.}
\vspace{-5pt}
\end{figure}

\subsection{The Generated Loss Functions}

In this section, we would like to show the generated loss functions and the loss generation process of our framework described in Section \ref{sec:process} and Figure \ref{fig:process}. We run the AutoLossGen framework under four model-dataset combinations for both the classification task and the regression task. 
The best loss function after Phase III under each setting is shown in Table \ref{Table:loss generation}, and more losses are provided in Appendix. The smoothing coefficient $\epsilon$ is included in each formulation when a division is calculated, which is to prevent numerical errors such as division by zero. The generated losses for the classification task are new losses and we name them MaxR, SumR and LogMin loss, respectively, while the generated loss for the regression task is the existing MSE loss for all of the four model-dataset combinations.

We take the generation process of the Max Ratio Loss (MaxR) as an example. Figure \ref{fig:process exp} shows the performance of the MF model on training and validation datasets during the loss search phase on the classification task. We can see that there is a significant increase on recommendation performance by using MaxR in that epoch, implying that MaxR is a good loss candidate to promote the performance. After the first phase, we confirm that MaxR is an effective loss based on the validation check in the second phase, which is further selected as the best loss in the third phase. And finally, we show that MaxR is indeed an effective loss on test set in Section \ref{sec:performance}.

Other generated loss functions also show similar improvements in Phase I and proven effective in Phase II, and finally selected by Phase III. Note that the generated loss under MF-Electronics setting is very similar with that under MLP-ML100K setting, with only a small difference in the constant term. Thus, we merge these two loss functions as SumR, whose name is due to the sum of the reciprocal of $\hat{y}$ and $y$. The SumR loss may not be handcrafted by human experts since if separating the formula to the sum of the reciprocals (i.e., $\frac{1}{\hat{y}}+\frac{1}{y}$), we intuitively may not accept it as a good loss since it may cause numerical exception. However, after merging the reciprocals and adding smoothing coefficient $\epsilon$, SumR proves to be a simple and effective loss, as we will show in the following experiment. For regression task, our AutoLossGen framework generates MSE as the loss in all of the four experimental settings, which is reasonable because MSE directly optimizes the target RMSE metric. This shows that when the ground-truth loss exists for a task, our framework is able to recover the ground-truth loss.

\subsection{Performance Comparison} 
\label{sec:performance}

\begin{figure}[t]
\vspace{-10pt}
  \centering
  \includegraphics[width=0.4\textwidth]{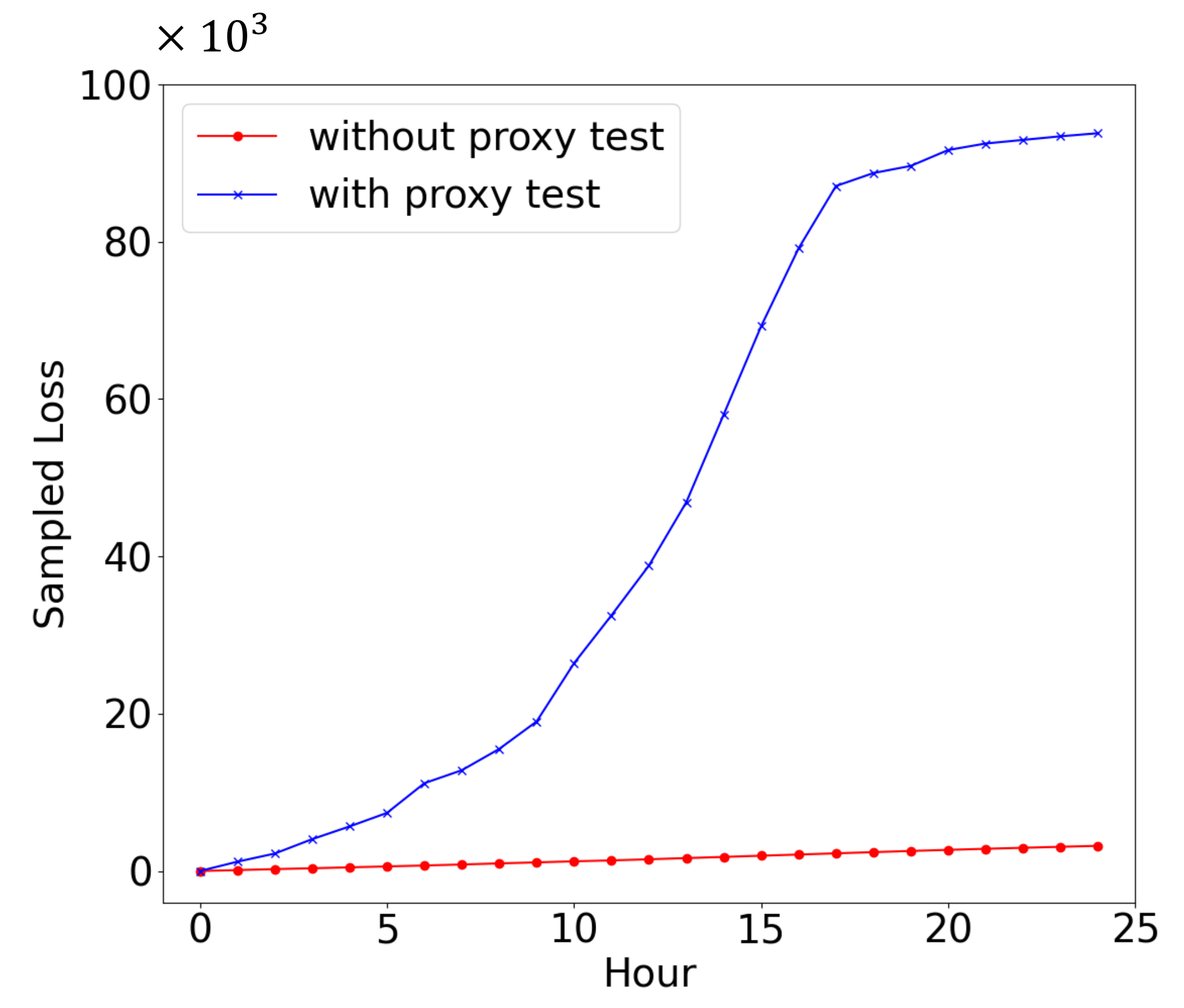}
  \vspace{-15pt}
  \caption{The accumulated number of sampled loss functions is shown in hours when using proxy test or not.}
  \label{fig:efficiency}
\end{figure}

\begin{table}[t]
    \centering
    \setlength{\tabcolsep}{1.7pt}
    \begin{tabular}{c|c|c|ccc|c}
        \toprule
        \multirow{2.5}{*}{Model} & \multirow{2.5}{*}{Dataset} & \multirow{2.5}{*}{Metric} & \multicolumn{3}{c|}{Handcrafted Loss} & Generated Loss \\
        \cmidrule{4-7}
        & & & BCE & Hinge & Focal & MSE
        \\
        \midrule \midrule
        \multirow{2}{*}{MF} & \multirow{2}{*}{ML-100K} & RMSE $\downarrow$ & 0.4540 & 0.4602 & 0.4710 & $0.4480^*$ \\ 
        & & MAE $\downarrow$ & 0.4386 & 0.4077 & 0.4673 & $0.3548^*$ \\
        \midrule
        \multirow{2}{*}{MF} & \multirow{2}{*}{Electronics} & RMSE $\downarrow$ & 0.4043 & 0.4075 & 0.4363  & $0.3945^*$ \\
        & & MAE $\downarrow$ & 0.3224 & 0.3501 & 0.4237 & $0.3180^*$ \\
        \midrule \midrule
        \multirow{2}{*}{MLP} & \multirow{2}{*}{ML-100K} & RMSE $\downarrow$ & 0.4338 & 0.4643 & 0.4376 & $0.4268^*$ \\
        & & MAE $\downarrow$ & 0.3802 & 0.3642 & 0.3811 & $0.3586^*$ \\
        \midrule
        \multirow{2}{*}{MLP} & \multirow{2}{*}{Electronics} & RMSE $\downarrow$ & 0.4005 & 0.4097 & 0.4153 & $0.4000^*$ \\ 
        & & MAE $\downarrow$ & 0.3105 & 0.3633 & 0.3573 & $0.3100^*$ \\
        \bottomrule
    \end{tabular}
    \vspace{5pt}
    \caption{Final performance on the regression task. $\downarrow$ means the measure is the lower the better. * represents the best performance for each row.}
    \vspace{-10pt}
    \label{Table:result on regression}
    \vspace{-10pt}
\end{table}

\begin{table}[t]
    \centering
    \begin{tabular}{l|r|r}
        \toprule
         & \#Sampled Loss & Speed-up \\
         \midrule
         Without proxy test & 3,200 & $1\times$\\
         With proxy test & 93,792 & $\sim30\times$\\
         \bottomrule
    \end{tabular}
    \vspace{5pt}
    \caption{Ablation study on proxy test. Speed-up shows how many times of losses are sampled than that w/o proxy test.}
    \vspace{-10pt}
    \label{Table:proxy test}
\end{table}

We compare the performance of the generated loss functions and four baseline losses on the test set for the classification task. Results under different model-dataset and evaluation metrics are shown in Table \ref{Table:result}. We have the following observations from the results.

First and most importantly, the loss function generated in the corresponding model-dataset performs better on AUC than any of the four baseline losses. The reason why the generated loss from our AutoLossGen framework outperforms the handcrafted loss functions is that during the loss generation process, the generated loss has been tested effective on both real data in Phase I (loss search) and on synthesized data in Phase II (validation check). Also, the performance on F1-score and accuracy of our generated losses is on par or better than that of all baselines. 
As a result, the generated losses from AutoLossGen can be more suitable for the corresponding model-dataset combination.

One interesting observation is that there is no globally best loss function, not only among the generated loss functions, but also for the handcrafted loss functions. For example, when we use MF as the RS model and Electronics as the dataset, hinge loss outperforms other baseline losses, however, if dataset switches to ML-100K, Focal loss is the best among the handcrafted losses. Furthermore, for the combination of MLP and ML-100K, BCE loss defeats other baseline losses on all three metrics. This observation indicates the importance of using AutoLossGen framework when the environment changes so as to find the best loss function tailored to the environment. 

For the regression task, the experimental results are show in Table \ref{Table:result on regression}. MSE loss is the generated loss in this task and the performance of MSE loss is better than other losses as expected.

\subsection{Loss Transferablity}

Even though the transferablity of the generated loss functions is not the key focus of this paper, we still do not expect the loss from AutoGenLoss can only be applicable to one model or one dataset. Table \ref{Table:result} also shows the results when a loss generated from one model-dataset setting is applied on another model-dataset setting.
We can see that even if applied to other experimental settings, our generated loss functions still outperform the baseline losses in most cases. The only exception is the LogMin loss under the MF-Electronics combination, where LogMin is slightly worse than the Hinge loss. However, LogMin is still better than all of the other three baseline losses under this setting. Besides, MaxR and SumR are both better than Hinge loss under this setting.

As a result, though it is best to apply the AutoLossGen framework to each specific experimental setting to obtain the most suitable loss for that setting, but to some extent, our generated loss functions are transferable to other experiment settings.

\subsection{Ablation Study on Efficiency}
\label{sec:ablation}

In this section, we discuss the improvement on efficiency by the proposed proxy test mechanism in Section \ref{sec:proxy test}.

For proxy test, we compare the number of explored loss functions with and without the proxy test mechanism under the same amount of time. Here, without proxy test mechanism means that all of the sampled loss from the controller will directly pass the proxy test and update the copied RS model.
For fair comparison, all experiments are run on a single NVIDIA Geforce 2080Ti GPU in 24 hours. The operating system is Ubuntu 20.04 LTS. The quantitative results are shown in Table \ref{Table:proxy test} and the accumulated number of sampled loss functions over time (in hours) is shown in Figure \ref{fig:efficiency}. We can see that the loss search efficiency is significantly better when the proxy test is applied, which means that a lot more loss functions can be explored within the same amount of time. This is not surprising because we encourage fewer restrictions on the form of sampled functions, which can lead to a large number of zero-gradient and gradient-level duplicated functions during the search process. Without the proxy test mechanism, all functions will pass the test and thus a lot of time has to be spent on updating the copied RS model, which reduces the number of functions that can be explored in the same amount of time.

\subsection{Ablation Study on Reward Filtering}
We explore the role of the reward filtering mechanism in Section \ref{sec:reward filtering}. We use the MF model under the ML-100K dataset as an example. Observations on other model-dataset combinations are similar.
We plot the accuracy of the RS model when the reward filtering mechanism is removed during the loss search phase as Figure \ref{fig:wandering} for better comparison with Figure \ref{fig:process exp}. We can see that when the reward filtering mechanism is removed, the AUC of the RS model will stay around 0.5, which means that the RS model is unable to learn useful information for prediction.

Intuitively, the reward filtering mechanism guarantees that the RS model will only be updated when the sampled loss is relatively good (i.e., reward $\ge-\eta$). Without the reward filtering mechanism, the RS model will always be updated by any sampled loss as long as the loss passes the proxy test, including those low-quality losses that lead to very negative rewards on the copied RS model.
As a result, the reward filtering mechanism is important to guarantee the performance of the RS model and to filter out the low-quality losses from the candidate loss list. 

\section{Conclusions and Future Work}
\label{sec:conclusions}

In this paper, we propose AutoGenLoss, an automatic loss function generation framework for recommender systems (RS), which is implemented by reinforcement learning (RL) and optimized in iterative and alternating schedules. 
Experiments show the better performance and transferablity of the generated loss functions than commonly used handcrafted loss functions under various settings.

We will further extend our framework on several aspects in the future. For the controller model, although REINFORCE \cite{williams1992simple} has shown its effectiveness, more state-of-the-art RL algorithms may reduce the redundant sampling for better efficiency. Meanwhile, faster search makes it possible to include more operators with larger search space to locate better loss functions. Another line of potential research is to propose an end-to-end differentiable model and integrate coefficient search in loss generation, since coefficients in loss functions may influence the performance. Besides, we mainly focused on classification and regression loss generation in this work, while it is promising to generalize the framework for other tasks such as ranking in the future. Finally, through this work we can see that the generated loss function formulas have different forms---some are simple while some are complex; some are effective while some are non-effective. In the future, it will be very interesting to build systematic theories and/or methodologies to understand what are the key factors that make a loss formula effective and how can such understanding be encoded into the loss learning algorithm to search for the best loss function more efficiently and effectively.

\section*{Acknowledgement}
This work was supported in part by NSF IIS 1910154, 2007907, and 2046457. Any opinions, findings, conclusions or recommendations expressed in this material are those of the authors and do not necessarily reflect those of the sponsors.

\begin{table*}[t]
    \centering
    \setlength{\tabcolsep}{5pt}
    \begin{tabular}{c|c|c}
        \toprule
        Positive Rate & Loss Formula & AUC (MF on ML-100K)\\
        \midrule
        1.000 & max$\big(\hat{y}, y * $max$(\hat{y}, \frac{1}{\hat{y} + \epsilon})\big)$ & 0.4882 \\
        1.000 & $\hat{y} + \frac{1}{y * \hat{y} + \epsilon}$ & 0.4918\\
        1.000 & $\frac{\text{max}(\hat{y}, y)}{\hat{y} + \epsilon}$ & 0.4881 \\
        1.000 & 1 + $\hat{y} + \frac{1}{\min(\hat{y} , y) + \epsilon} + \max\Big( \frac{1}{\min(\hat{y} , y) + \epsilon} , \max\big(\min(\hat{y} , y) , \hat{y} + \frac{1}{\min(\hat{y} , y) + \epsilon}  \big) \Big) $  & 0.4863\\
        1.000 & $\hat{y} + y + \frac{1}{\text{min}(y, \hat{y})^2 + \epsilon}$ & 0.4898 \\
        1.000 & $\frac{1}{\text{min}(y, \hat{y}) + \epsilon} + \text{max}(\hat{y}, y) + \text{max}(\hat{y}, y)^2$ & 0.4877 \\
        1.000 & $1 + \text{min}\big(\text{min}(1, \hat{y} + y), \frac{1}{y + \hat{y} + \epsilon}\big)$ & \textbf{0.7658} \\
        1.000 & max$(y, \hat{y}) + \frac{1}{\text{max}(\hat{y} + 1, y) + \epsilon}$ & \textbf{0.7703}\\
        1.000 & $\hat{y} + \frac{1}{\text{min}(\hat{y}, y) + \epsilon}$ & 0.4872 \\
        1.000 & $\Big(\text{min}\big(\frac{1}{y + \hat{y} + \epsilon}, \text{max}(y, \hat{y})\big)\Big)^2$ & \textbf{0.7782} \\
        0.996 & $\min\Big(\frac{1}{\text{max}(\hat{y}, y + \hat{y}^2) + \epsilon}, \text{max}\big(y, (\hat{y}^2)^2\big)\Big)$ & \textbf{0.7787}\\
        0.902 & $\hat{y}^2 + \frac{1}{\text{min}(y, \hat{y}) + \epsilon}$ & 0.4884\\
        \bottomrule
    \end{tabular}
    \caption{Loss generation result, changing random seed from 0 to 42. $\epsilon$ is a tunable parameter representing the smoothing coefficient to prevent numerical errors such as division by zero. Filter out the loss functions whose positive rate is less than 0.9 (i.e., randomly sample 2000 pairs of $(y, \hat{y})$ and fewer than 1800 pairs lead $\hat{y}$ towards $y$) in Validation Check.}
    \label{Table:loss generation different random seed}
    \vspace{-20pt}
\end{table*}

\section*{Appendix}
In the AutoLossGen framework, multiple loss functions may pass Phase II (validation check) and enter Phase III (effectiveness test). In Table \ref{Table:loss generation} we have listed the best loss $f^*$ after Phase III (effectiveness test) on each dataset-model combination. Here, we use the MF on ML-100K combination as an example to show all of the generated loss functions that passed Phase II (i.e., passed the validation check on more than 90\% of $(y, \hat{y})$ pairs) and entered Phase III, as shown in Table \ref{Table:loss generation different random seed}. An observation from the results is that the generated loss is either effective (AUC close to known loss functions) or non-effective at all (AUC around 0.5, i.e., close to random guess). This implies that the effectiveness (in terms of AUC) of loss functions does not uniformly span across the effectiveness space, but instead tend to be binary, i.e., a loss function either works or does not work at all, and there may not be a loss function that partially works. As long as an effective loss exists, the AutoLossGen framework is able to generate the loss in Phase I and eventually find it through Phase II (validation check) and Phase III (effectivenss test). Our observation also indicates that there may exist some general knowledge about what key factors contribute to the effectiveness of loss formula. For example, the best loss functions (Table \ref{Table:loss generation}) tend to exhibit certain degrees of symmetry. In the future, it will be very interesting to build systematic theories and/or methodologies to understand what are the key factors that make a loss formula effective and how can such knowledge be embedded into the loss learning algorithm or process as prior knowledge to search for the best loss function more efficiently and effectively.

\bibliographystyle{ACM-Reference-Format}
\bibliography{paper.bib}


\begin{thebibliography}{60}


\ifx \showCODEN    \undefined \def \showCODEN     #1{\unskip}     \fi
\ifx \showDOI      \undefined \def \showDOI       #1{#1}\fi
\ifx \showISBNx    \undefined \def \showISBNx     #1{\unskip}     \fi
\ifx \showISBNxiii \undefined \def \showISBNxiii  #1{\unskip}     \fi
\ifx \showISSN     \undefined \def \showISSN      #1{\unskip}     \fi
\ifx \showLCCN     \undefined \def \showLCCN      #1{\unskip}     \fi
\ifx \shownote     \undefined \def \shownote      #1{#1}          \fi
\ifx \showarticletitle \undefined \def \showarticletitle #1{#1}   \fi
\ifx \showURL      \undefined \def \showURL       {\relax}        \fi
\providecommand\bibfield[2]{#2}
\providecommand\bibinfo[2]{#2}
\providecommand\natexlab[1]{#1}
\providecommand\showeprint[2][]{arXiv:#2}

\bibitem[\protect\citeauthoryear{Allen}{Allen}{1971}]%
        {allen1971mean}
\bibfield{author}{\bibinfo{person}{David~M Allen}.}
  \bibinfo{year}{1971}\natexlab{}.
\newblock \showarticletitle{Mean square error of prediction as a criterion for
  selecting variables}.
\newblock \bibinfo{journal}{\emph{Technometrics}} \bibinfo{volume}{13},
  \bibinfo{number}{3} (\bibinfo{year}{1971}), \bibinfo{pages}{469--475}.
\newblock


\bibitem[\protect\citeauthoryear{Anandalingam and Friesz}{Anandalingam and
  Friesz}{1992}]%
        {anandalingam1992hierarchical}
\bibfield{author}{\bibinfo{person}{G Anandalingam} {and}
  \bibinfo{person}{Terry~L Friesz}.} \bibinfo{year}{1992}\natexlab{}.
\newblock \showarticletitle{Hierarchical optimization: An introduction}.
\newblock \bibinfo{journal}{\emph{Annals of Operations Research}}
  \bibinfo{volume}{34}, \bibinfo{number}{1} (\bibinfo{year}{1992}),
  \bibinfo{pages}{1--11}.
\newblock


\bibitem[\protect\citeauthoryear{Baker, Gupta, Naik, and Raskar}{Baker
  et~al\mbox{.}}{2017}]%
        {baker2016designing}
\bibfield{author}{\bibinfo{person}{Bowen Baker}, \bibinfo{person}{Otkrist
  Gupta}, \bibinfo{person}{Nikhil Naik}, {and} \bibinfo{person}{Ramesh
  Raskar}.} \bibinfo{year}{2017}\natexlab{}.
\newblock \showarticletitle{Designing Neural Network Architectures using
  Reinforcement Learning}. In \bibinfo{booktitle}{\emph{5th International
  Conference on Learning Representations, {ICLR} 2017, Toulon, France, April
  24-26, 2017, Conference Track Proceedings}}.
  \bibinfo{publisher}{OpenReview.net}.
\newblock
\urldef\tempurl%
\url{https://openreview.net/forum?id=S1c2cvqee}
\showURL{%
\tempurl}


\bibitem[\protect\citeauthoryear{Bello, Pham, Le, Norouzi, and Bengio}{Bello
  et~al\mbox{.}}{2016}]%
        {bello2016neural}
\bibfield{author}{\bibinfo{person}{Irwan Bello}, \bibinfo{person}{Hieu Pham},
  \bibinfo{person}{Quoc~V Le}, \bibinfo{person}{Mohammad Norouzi}, {and}
  \bibinfo{person}{Samy Bengio}.} \bibinfo{year}{2016}\natexlab{}.
\newblock \showarticletitle{Neural combinatorial optimization with
  reinforcement learning}.
\newblock \bibinfo{journal}{\emph{arXiv preprint arXiv:1611.09940}}
  (\bibinfo{year}{2016}).
\newblock


\bibitem[\protect\citeauthoryear{Bello, Zoph, Vasudevan, and Le}{Bello
  et~al\mbox{.}}{2017}]%
        {bello2017neural}
\bibfield{author}{\bibinfo{person}{Irwan Bello}, \bibinfo{person}{Barret Zoph},
  \bibinfo{person}{Vijay Vasudevan}, {and} \bibinfo{person}{Quoc~V Le}.}
  \bibinfo{year}{2017}\natexlab{}.
\newblock \showarticletitle{Neural optimizer search with reinforcement
  learning}. In \bibinfo{booktitle}{\emph{International Conference on Machine
  Learning}}. PMLR, \bibinfo{pages}{459--468}.
\newblock


\bibitem[\protect\citeauthoryear{Cai, Yang, Zhang, Han, and Yu}{Cai
  et~al\mbox{.}}{2018}]%
        {cai2018path}
\bibfield{author}{\bibinfo{person}{Han Cai}, \bibinfo{person}{Jiacheng Yang},
  \bibinfo{person}{Weinan Zhang}, \bibinfo{person}{Song Han}, {and}
  \bibinfo{person}{Yong Yu}.} \bibinfo{year}{2018}\natexlab{}.
\newblock \showarticletitle{Path-level network transformation for efficient
  architecture search}. In \bibinfo{booktitle}{\emph{International Conference
  on Machine Learning}}. PMLR, \bibinfo{pages}{678--687}.
\newblock


\bibitem[\protect\citeauthoryear{Calders and Jaroszewicz}{Calders and
  Jaroszewicz}{2007}]%
        {calders2007efficient}
\bibfield{author}{\bibinfo{person}{Toon Calders} {and} \bibinfo{person}{Szymon
  Jaroszewicz}.} \bibinfo{year}{2007}\natexlab{}.
\newblock \showarticletitle{Efficient AUC optimization for classification}. In
  \bibinfo{booktitle}{\emph{European Conference on Principles of Data Mining
  and Knowledge Discovery}}. Springer, \bibinfo{pages}{42--53}.
\newblock


\bibitem[\protect\citeauthoryear{Cerqueira, Moniz, and Soares}{Cerqueira
  et~al\mbox{.}}{2021}]%
        {cerqueira2021vest}
\bibfield{author}{\bibinfo{person}{Vitor Cerqueira}, \bibinfo{person}{Nuno
  Moniz}, {and} \bibinfo{person}{Carlos Soares}.}
  \bibinfo{year}{2021}\natexlab{}.
\newblock \showarticletitle{Vest: Automatic feature engineering for
  forecasting}.
\newblock \bibinfo{journal}{\emph{Machine Learning}} (\bibinfo{year}{2021}),
  \bibinfo{pages}{1--23}.
\newblock


\bibitem[\protect\citeauthoryear{Chai and Draxler}{Chai and Draxler}{2014}]%
        {chai2014root}
\bibfield{author}{\bibinfo{person}{Tianfeng Chai} {and}
  \bibinfo{person}{Roland~R Draxler}.} \bibinfo{year}{2014}\natexlab{}.
\newblock \showarticletitle{Root mean square error (RMSE) or mean absolute
  error (MAE)?--Arguments against avoiding RMSE in the literature}.
\newblock \bibinfo{journal}{\emph{Geoscientific model development}}
  \bibinfo{volume}{7}, \bibinfo{number}{3} (\bibinfo{year}{2014}),
  \bibinfo{pages}{1247--1250}.
\newblock


\bibitem[\protect\citeauthoryear{Charbonnier, Blanc-Feraud, Aubert, and
  Barlaud}{Charbonnier et~al\mbox{.}}{1994}]%
        {charbonnier1994two}
\bibfield{author}{\bibinfo{person}{Pierre Charbonnier}, \bibinfo{person}{Laure
  Blanc-Feraud}, \bibinfo{person}{Gilles Aubert}, {and} \bibinfo{person}{Michel
  Barlaud}.} \bibinfo{year}{1994}\natexlab{}.
\newblock \showarticletitle{Two deterministic half-quadratic regularization
  algorithms for computed imaging}. In \bibinfo{booktitle}{\emph{Proceedings of
  1st International Conference on Image Processing}}, Vol.~\bibinfo{volume}{2}.
  IEEE, \bibinfo{pages}{168--172}.
\newblock


\bibitem[\protect\citeauthoryear{Chen, Li, Shi, Liu, Zhu, and Zhang}{Chen
  et~al\mbox{.}}{2022}]%
        {chen2022graph}
\bibfield{author}{\bibinfo{person}{Hanxiong Chen}, \bibinfo{person}{Yunqi Li},
  \bibinfo{person}{Shaoyun Shi}, \bibinfo{person}{Shuchang Liu},
  \bibinfo{person}{He Zhu}, {and} \bibinfo{person}{Yongfeng Zhang}.}
  \bibinfo{year}{2022}\natexlab{}.
\newblock \showarticletitle{Graph Collaborative Reasoning}. In
  \bibinfo{booktitle}{\emph{Proceedings of the Fifteenth ACM International
  Conference on Web Search and Data Mining}}. \bibinfo{pages}{75--84}.
\newblock


\bibitem[\protect\citeauthoryear{Chen, Shi, Li, and Zhang}{Chen
  et~al\mbox{.}}{2021}]%
        {chen2021neural}
\bibfield{author}{\bibinfo{person}{Hanxiong Chen}, \bibinfo{person}{Shaoyun
  Shi}, \bibinfo{person}{Yunqi Li}, {and} \bibinfo{person}{Yongfeng Zhang}.}
  \bibinfo{year}{2021}\natexlab{}.
\newblock \showarticletitle{Neural Collaborative Reasoning}. In
  \bibinfo{booktitle}{\emph{Proceedings of the Web Conference 2021}}.
  \bibinfo{pages}{1516--1527}.
\newblock


\bibitem[\protect\citeauthoryear{Cheng, Koc, Harmsen, Shaked, Chandra, Aradhye,
  Anderson, Corrado, Chai, Ispir, et~al\mbox{.}}{Cheng et~al\mbox{.}}{2016}]%
        {cheng2016wide}
\bibfield{author}{\bibinfo{person}{Heng-Tze Cheng}, \bibinfo{person}{Levent
  Koc}, \bibinfo{person}{Jeremiah Harmsen}, \bibinfo{person}{Tal Shaked},
  \bibinfo{person}{Tushar Chandra}, \bibinfo{person}{Hrishi Aradhye},
  \bibinfo{person}{Glen Anderson}, \bibinfo{person}{Greg Corrado},
  \bibinfo{person}{Wei Chai}, \bibinfo{person}{Mustafa Ispir}, {et~al\mbox{.}}}
  \bibinfo{year}{2016}\natexlab{}.
\newblock \showarticletitle{Wide \& deep learning for recommender systems}. In
  \bibinfo{booktitle}{\emph{Proceedings of the 1st workshop on deep learning
  for recommender systems}}. \bibinfo{pages}{7--10}.
\newblock


\bibitem[\protect\citeauthoryear{Feurer, Klein, Eggensperger, Springenberg,
  Blum, and Hutter}{Feurer et~al\mbox{.}}{2015}]%
        {NIPS2015_11d0e628}
\bibfield{author}{\bibinfo{person}{Matthias Feurer}, \bibinfo{person}{Aaron
  Klein}, \bibinfo{person}{Katharina Eggensperger}, \bibinfo{person}{Jost
  Springenberg}, \bibinfo{person}{Manuel Blum}, {and} \bibinfo{person}{Frank
  Hutter}.} \bibinfo{year}{2015}\natexlab{}.
\newblock \showarticletitle{Efficient and Robust Automated Machine Learning}.
  In \bibinfo{booktitle}{\emph{Advances in Neural Information Processing
  Systems}}, \bibfield{editor}{\bibinfo{person}{C.~Cortes},
  \bibinfo{person}{N.~Lawrence}, \bibinfo{person}{D.~Lee},
  \bibinfo{person}{M.~Sugiyama}, {and} \bibinfo{person}{R.~Garnett}} (Eds.),
  Vol.~\bibinfo{volume}{28}. \bibinfo{publisher}{Curran Associates, Inc.}
\newblock
\urldef\tempurl%
\url{https://proceedings.neurips.cc/paper/2015/file/11d0e6287202fced83f79975ec59a3a6-Paper.pdf}
\showURL{%
\tempurl}


\bibitem[\protect\citeauthoryear{Fu, Xian, Zhu, Xu, Li, De~Melo, and Zhang}{Fu
  et~al\mbox{.}}{2021}]%
        {fu2021hoops}
\bibfield{author}{\bibinfo{person}{Zuohui Fu}, \bibinfo{person}{Yikun Xian},
  \bibinfo{person}{Yaxin Zhu}, \bibinfo{person}{Shuyuan Xu},
  \bibinfo{person}{Zelong Li}, \bibinfo{person}{Gerard De~Melo}, {and}
  \bibinfo{person}{Yongfeng Zhang}.} \bibinfo{year}{2021}\natexlab{}.
\newblock \showarticletitle{HOOPS: Human-in-the-Loop Graph Reasoning for
  Conversational Recommendation}. In \bibinfo{booktitle}{\emph{Proceedings of
  the 44th International ACM SIGIR Conference on Research and Development in
  Information Retrieval}}. \bibinfo{pages}{2415--2421}.
\newblock


\bibitem[\protect\citeauthoryear{Ge, Tan, Zhu, Xia, Luo, Liu, Fu, Geng, Li, and
  Zhang}{Ge et~al\mbox{.}}{2022}]%
        {ge2022explainable}
\bibfield{author}{\bibinfo{person}{Yingqiang Ge}, \bibinfo{person}{Juntao Tan},
  \bibinfo{person}{Yan Zhu}, \bibinfo{person}{Yinglong Xia},
  \bibinfo{person}{Jiebo Luo}, \bibinfo{person}{Shuchang Liu},
  \bibinfo{person}{Zuohui Fu}, \bibinfo{person}{Shijie Geng},
  \bibinfo{person}{Zelong Li}, {and} \bibinfo{person}{Yongfeng Zhang}.}
  \bibinfo{year}{2022}\natexlab{}.
\newblock \showarticletitle{Explainable Fairness in Recommendation}.
\newblock \bibinfo{journal}{\emph{SIGIR}} (\bibinfo{year}{2022}).
\newblock


\bibitem[\protect\citeauthoryear{Ge, Xu, Liu, Fu, Sun, and Zhang}{Ge
  et~al\mbox{.}}{2020}]%
        {ge2020learning}
\bibfield{author}{\bibinfo{person}{Yingqiang Ge}, \bibinfo{person}{Shuyuan Xu},
  \bibinfo{person}{Shuchang Liu}, \bibinfo{person}{Zuohui Fu},
  \bibinfo{person}{Fei Sun}, {and} \bibinfo{person}{Yongfeng Zhang}.}
  \bibinfo{year}{2020}\natexlab{}.
\newblock \showarticletitle{Learning Personalized Risk Preferences for
  Recommendation}. In \bibinfo{booktitle}{\emph{Proceedings of the 43rd
  SIGIR}}. \bibinfo{pages}{409--418}.
\newblock


\bibitem[\protect\citeauthoryear{Ge, Xu, Liu, Geng, Fu, and Zhang}{Ge
  et~al\mbox{.}}{2019}]%
        {ge2019maximizing}
\bibfield{author}{\bibinfo{person}{Yingqiang Ge}, \bibinfo{person}{Shuyuan Xu},
  \bibinfo{person}{Shuchang Liu}, \bibinfo{person}{Shijie Geng},
  \bibinfo{person}{Zuohui Fu}, {and} \bibinfo{person}{Yongfeng Zhang}.}
  \bibinfo{year}{2019}\natexlab{}.
\newblock \showarticletitle{Maximizing marginal utility per dollar for economic
  recommendation}. In \bibinfo{booktitle}{\emph{The World Wide Web
  Conference}}. \bibinfo{pages}{2757--2763}.
\newblock


\bibitem[\protect\citeauthoryear{Geng, Liu, Fu, Ge, and Zhang}{Geng
  et~al\mbox{.}}{2022}]%
        {geng2022recommendation}
\bibfield{author}{\bibinfo{person}{Shijie Geng}, \bibinfo{person}{Shuchang
  Liu}, \bibinfo{person}{Zuohui Fu}, \bibinfo{person}{Yingqiang Ge}, {and}
  \bibinfo{person}{Yongfeng Zhang}.} \bibinfo{year}{2022}\natexlab{}.
\newblock \showarticletitle{Recommendation as Language Processing (RLP): A
  Unified Pretrain, Personalized Prompt \& Predict Paradigm (P5)}.
\newblock \bibinfo{journal}{\emph{arXiv preprint arXiv:2203.13366}}
  (\bibinfo{year}{2022}).
\newblock


\bibitem[\protect\citeauthoryear{Girshick}{Girshick}{2015}]%
        {girshick2015fast}
\bibfield{author}{\bibinfo{person}{Ross Girshick}.}
  \bibinfo{year}{2015}\natexlab{}.
\newblock \showarticletitle{Fast r-cnn}. In
  \bibinfo{booktitle}{\emph{Proceedings of the IEEE international conference on
  computer vision}}. \bibinfo{pages}{1440--1448}.
\newblock


\bibitem[\protect\citeauthoryear{Hariharan, Zelnik-Manor, Vishwanathan, and
  Varma}{Hariharan et~al\mbox{.}}{2010}]%
        {hariharan2010large}
\bibfield{author}{\bibinfo{person}{Bharath Hariharan}, \bibinfo{person}{Lihi
  Zelnik-Manor}, \bibinfo{person}{SVN Vishwanathan}, {and}
  \bibinfo{person}{Manik Varma}.} \bibinfo{year}{2010}\natexlab{}.
\newblock \showarticletitle{Large scale max-margin multi-label classification
  with priors}. In \bibinfo{booktitle}{\emph{ICML}}.
\newblock


\bibitem[\protect\citeauthoryear{Harper and Konstan}{Harper and
  Konstan}{2015}]%
        {harper2015movielens}
\bibfield{author}{\bibinfo{person}{F~Maxwell Harper} {and}
  \bibinfo{person}{Joseph~A Konstan}.} \bibinfo{year}{2015}\natexlab{}.
\newblock \showarticletitle{The movielens datasets: History and context}.
\newblock \bibinfo{journal}{\emph{Acm transactions on interactive intelligent
  systems (tiis)}} \bibinfo{volume}{5}, \bibinfo{number}{4}
  (\bibinfo{year}{2015}), \bibinfo{pages}{1--19}.
\newblock


\bibitem[\protect\citeauthoryear{He and McAuley}{He and McAuley}{2016}]%
        {he2016ups}
\bibfield{author}{\bibinfo{person}{Ruining He} {and} \bibinfo{person}{Julian
  McAuley}.} \bibinfo{year}{2016}\natexlab{}.
\newblock \showarticletitle{Ups and downs: Modeling the visual evolution of
  fashion trends with one-class collaborative filtering}. In
  \bibinfo{booktitle}{\emph{proceedings of the 25th international conference on
  world wide web}}. \bibinfo{pages}{507--517}.
\newblock


\bibitem[\protect\citeauthoryear{Huang, Li, Cheng, Zhang, and Hauptmann}{Huang
  et~al\mbox{.}}{2018}]%
        {huang2018gnas}
\bibfield{author}{\bibinfo{person}{Siyu Huang}, \bibinfo{person}{Xi Li},
  \bibinfo{person}{Zhi-Qi Cheng}, \bibinfo{person}{Zhongfei Zhang}, {and}
  \bibinfo{person}{Alexander Hauptmann}.} \bibinfo{year}{2018}\natexlab{}.
\newblock \showarticletitle{Gnas: A greedy neural architecture search method
  for multi-attribute learning}. In \bibinfo{booktitle}{\emph{Proceedings of
  the 26th ACM international conference on Multimedia}}.
  \bibinfo{pages}{2049--2057}.
\newblock


\bibitem[\protect\citeauthoryear{Huber}{Huber}{1992}]%
        {huber1992robust}
\bibfield{author}{\bibinfo{person}{Peter~J Huber}.}
  \bibinfo{year}{1992}\natexlab{}.
\newblock \showarticletitle{Robust estimation of a location parameter}.
\newblock In \bibinfo{booktitle}{\emph{Breakthroughs in statistics}}.
  \bibinfo{publisher}{Springer}, \bibinfo{pages}{492--518}.
\newblock


\bibitem[\protect\citeauthoryear{Kanter and Veeramachaneni}{Kanter and
  Veeramachaneni}{2015}]%
        {kanter2015deep}
\bibfield{author}{\bibinfo{person}{James~Max Kanter} {and}
  \bibinfo{person}{Kalyan Veeramachaneni}.} \bibinfo{year}{2015}\natexlab{}.
\newblock \showarticletitle{Deep feature synthesis: Towards automating data
  science endeavors}. In \bibinfo{booktitle}{\emph{2015 IEEE international
  conference on data science and advanced analytics (DSAA)}}. IEEE,
  \bibinfo{pages}{1--10}.
\newblock


\bibitem[\protect\citeauthoryear{Katz, Shin, and Song}{Katz
  et~al\mbox{.}}{2016}]%
        {katz2016explorekit}
\bibfield{author}{\bibinfo{person}{Gilad Katz}, \bibinfo{person}{Eui
  Chul~Richard Shin}, {and} \bibinfo{person}{Dawn Song}.}
  \bibinfo{year}{2016}\natexlab{}.
\newblock \showarticletitle{Explorekit: Automatic feature generation and
  selection}. In \bibinfo{booktitle}{\emph{2016 IEEE 16th International
  Conference on Data Mining (ICDM)}}. IEEE, \bibinfo{pages}{979--984}.
\newblock


\bibitem[\protect\citeauthoryear{Koren, Bell, and Volinsky}{Koren
  et~al\mbox{.}}{2009}]%
        {koren2009matrix}
\bibfield{author}{\bibinfo{person}{Yehuda Koren}, \bibinfo{person}{Robert
  Bell}, {and} \bibinfo{person}{Chris Volinsky}.}
  \bibinfo{year}{2009}\natexlab{}.
\newblock \showarticletitle{Matrix factorization techniques for recommender
  systems}.
\newblock \bibinfo{journal}{\emph{Computer}} \bibinfo{volume}{42},
  \bibinfo{number}{8} (\bibinfo{year}{2009}), \bibinfo{pages}{30--37}.
\newblock


\bibitem[\protect\citeauthoryear{Kotthoff, Thornton, Hoos, Hutter, and
  Leyton-Brown}{Kotthoff et~al\mbox{.}}{2019}]%
        {kotthoff2019auto}
\bibfield{author}{\bibinfo{person}{Lars Kotthoff}, \bibinfo{person}{Chris
  Thornton}, \bibinfo{person}{Holger~H Hoos}, \bibinfo{person}{Frank Hutter},
  {and} \bibinfo{person}{Kevin Leyton-Brown}.} \bibinfo{year}{2019}\natexlab{}.
\newblock \showarticletitle{Auto-WEKA: Automatic model selection and
  hyperparameter optimization in WEKA}.
\newblock In \bibinfo{booktitle}{\emph{Automated Machine Learning}}.
  \bibinfo{publisher}{Springer, Cham}, \bibinfo{pages}{81--95}.
\newblock


\bibitem[\protect\citeauthoryear{Kullback and Leibler}{Kullback and
  Leibler}{1951}]%
        {kullback1951information}
\bibfield{author}{\bibinfo{person}{Solomon Kullback} {and}
  \bibinfo{person}{Richard~A Leibler}.} \bibinfo{year}{1951}\natexlab{}.
\newblock \showarticletitle{On information and sufficiency}.
\newblock \bibinfo{journal}{\emph{The annals of mathematical statistics}}
  \bibinfo{volume}{22}, \bibinfo{number}{1} (\bibinfo{year}{1951}),
  \bibinfo{pages}{79--86}.
\newblock


\bibitem[\protect\citeauthoryear{Lee and Lin}{Lee and Lin}{2013}]%
        {lee2013study}
\bibfield{author}{\bibinfo{person}{Ching-Pei Lee} {and}
  \bibinfo{person}{Chih-Jen Lin}.} \bibinfo{year}{2013}\natexlab{}.
\newblock \showarticletitle{A study on L2-loss (squared hinge-loss) multiclass
  SVM}.
\newblock \bibinfo{journal}{\emph{Neural computation}} \bibinfo{volume}{25},
  \bibinfo{number}{5} (\bibinfo{year}{2013}), \bibinfo{pages}{1302--1323}.
\newblock


\bibitem[\protect\citeauthoryear{Li, Yuan, Lin, Guo, Wu, Yan, and Ouyang}{Li
  et~al\mbox{.}}{2019}]%
        {li2019lfs}
\bibfield{author}{\bibinfo{person}{Chuming Li}, \bibinfo{person}{Xin Yuan},
  \bibinfo{person}{Chen Lin}, \bibinfo{person}{Minghao Guo},
  \bibinfo{person}{Wei Wu}, \bibinfo{person}{Junjie Yan}, {and}
  \bibinfo{person}{Wanli Ouyang}.} \bibinfo{year}{2019}\natexlab{}.
\newblock \showarticletitle{Am-lfs: Automl for loss function search}. In
  \bibinfo{booktitle}{\emph{Proceedings of the IEEE/CVF International
  Conference on Computer Vision}}. \bibinfo{pages}{8410--8419}.
\newblock


\bibitem[\protect\citeauthoryear{Li, Fu, Dai, Li, Huang, and Zhu}{Li
  et~al\mbox{.}}{2021a}]%
        {li2021autoloss}
\bibfield{author}{\bibinfo{person}{Hao Li}, \bibinfo{person}{Tianwen Fu},
  \bibinfo{person}{Jifeng Dai}, \bibinfo{person}{Hongsheng Li},
  \bibinfo{person}{Gao Huang}, {and} \bibinfo{person}{Xizhou Zhu}.}
  \bibinfo{year}{2021}\natexlab{a}.
\newblock \showarticletitle{AutoLoss-Zero: Searching Loss Functions from
  Scratch for Generic Tasks}.
\newblock \bibinfo{journal}{\emph{arXiv preprint arXiv:2103.14026}}
  (\bibinfo{year}{2021}).
\newblock


\bibitem[\protect\citeauthoryear{Li, Tao, Zhu, Wang, Huang, and Dai}{Li
  et~al\mbox{.}}{2021b}]%
        {li2020auto}
\bibfield{author}{\bibinfo{person}{Hao Li}, \bibinfo{person}{Chenxin Tao},
  \bibinfo{person}{Xizhou Zhu}, \bibinfo{person}{Xiaogang Wang},
  \bibinfo{person}{Gao Huang}, {and} \bibinfo{person}{Jifeng Dai}.}
  \bibinfo{year}{2021}\natexlab{b}.
\newblock \showarticletitle{Auto Seg-Loss: Searching Metric Surrogates for
  Semantic Segmentation}. In \bibinfo{booktitle}{\emph{International Conference
  on Learning Representations}}.
\newblock
\urldef\tempurl%
\url{https://openreview.net/forum?id=MJAqnaC2vO1}
\showURL{%
\tempurl}


\bibitem[\protect\citeauthoryear{Lin, Goyal, Girshick, He, and Doll{\'a}r}{Lin
  et~al\mbox{.}}{2017}]%
        {lin2017focal}
\bibfield{author}{\bibinfo{person}{Tsung-Yi Lin}, \bibinfo{person}{Priya
  Goyal}, \bibinfo{person}{Ross Girshick}, \bibinfo{person}{Kaiming He}, {and}
  \bibinfo{person}{Piotr Doll{\'a}r}.} \bibinfo{year}{2017}\natexlab{}.
\newblock \showarticletitle{Focal loss for dense object detection}. In
  \bibinfo{booktitle}{\emph{Proceedings of the IEEE international conference on
  computer vision}}. \bibinfo{pages}{2980--2988}.
\newblock


\bibitem[\protect\citeauthoryear{Liu, Zoph, Neumann, Shlens, Hua, Li, Fei-Fei,
  Yuille, Huang, and Murphy}{Liu et~al\mbox{.}}{2018}]%
        {liu2018progressive}
\bibfield{author}{\bibinfo{person}{Chenxi Liu}, \bibinfo{person}{Barret Zoph},
  \bibinfo{person}{Maxim Neumann}, \bibinfo{person}{Jonathon Shlens},
  \bibinfo{person}{Wei Hua}, \bibinfo{person}{Li-Jia Li}, \bibinfo{person}{Li
  Fei-Fei}, \bibinfo{person}{Alan Yuille}, \bibinfo{person}{Jonathan Huang},
  {and} \bibinfo{person}{Kevin Murphy}.} \bibinfo{year}{2018}\natexlab{}.
\newblock \showarticletitle{Progressive neural architecture search}. In
  \bibinfo{booktitle}{\emph{Proceedings of the European conference on computer
  vision (ECCV)}}. \bibinfo{pages}{19--34}.
\newblock


\bibitem[\protect\citeauthoryear{Liu, Simonyan, and Yang}{Liu
  et~al\mbox{.}}{2019}]%
        {liu2018darts}
\bibfield{author}{\bibinfo{person}{Hanxiao Liu}, \bibinfo{person}{Karen
  Simonyan}, {and} \bibinfo{person}{Yiming Yang}.}
  \bibinfo{year}{2019}\natexlab{}.
\newblock \showarticletitle{{DARTS}: Differentiable Architecture Search}. In
  \bibinfo{booktitle}{\emph{International Conference on Learning
  Representations}}.
\newblock
\urldef\tempurl%
\url{https://openreview.net/forum?id=S1eYHoC5FX}
\showURL{%
\tempurl}


\bibitem[\protect\citeauthoryear{Liu, Zhang, Wang, Xu, Liang, Jiang, and
  Li}{Liu et~al\mbox{.}}{2021}]%
        {liu2021loss}
\bibfield{author}{\bibinfo{person}{Peidong Liu}, \bibinfo{person}{Gengwei
  Zhang}, \bibinfo{person}{Bochao Wang}, \bibinfo{person}{Hang Xu},
  \bibinfo{person}{Xiaodan Liang}, \bibinfo{person}{Yong Jiang}, {and}
  \bibinfo{person}{Zhenguo Li}.} \bibinfo{year}{2021}\natexlab{}.
\newblock \showarticletitle{Loss Function Discovery for Object Detection via
  Convergence-Simulation Driven Search}. In
  \bibinfo{booktitle}{\emph{International Conference on Learning
  Representations}}.
\newblock
\urldef\tempurl%
\url{https://openreview.net/forum?id=5jzlpHvvRk}
\showURL{%
\tempurl}


\bibitem[\protect\citeauthoryear{Liu and Lai}{Liu and Lai}{2020}]%
        {liu2020stochastic}
\bibfield{author}{\bibinfo{person}{Qingliang Liu} {and} \bibinfo{person}{Jinmei
  Lai}.} \bibinfo{year}{2020}\natexlab{}.
\newblock \showarticletitle{Stochastic Loss Function}.
\newblock \bibinfo{journal}{\emph{Proceedings of the AAAI Conference on
  Artificial Intelligence}} \bibinfo{volume}{34}, \bibinfo{number}{04}
  (\bibinfo{date}{Apr.} \bibinfo{year}{2020}), \bibinfo{pages}{4884--4891}.
\newblock
\urldef\tempurl%
\url{https://doi.org/10.1609/aaai.v34i04.5925}
\showDOI{\tempurl}


\bibitem[\protect\citeauthoryear{Liu, Wen, Yu, Li, Raj, and Song}{Liu
  et~al\mbox{.}}{2017}]%
        {liu2017sphereface}
\bibfield{author}{\bibinfo{person}{Weiyang Liu}, \bibinfo{person}{Yandong Wen},
  \bibinfo{person}{Zhiding Yu}, \bibinfo{person}{Ming Li},
  \bibinfo{person}{Bhiksha Raj}, {and} \bibinfo{person}{Le Song}.}
  \bibinfo{year}{2017}\natexlab{}.
\newblock \showarticletitle{Sphereface: Deep hypersphere embedding for face
  recognition}. In \bibinfo{booktitle}{\emph{Proceedings of the IEEE conference
  on computer vision and pattern recognition}}. \bibinfo{pages}{212--220}.
\newblock


\bibitem[\protect\citeauthoryear{Liu, Wen, Yu, and Yang}{Liu
  et~al\mbox{.}}{2016}]%
        {liu2016large}
\bibfield{author}{\bibinfo{person}{Weiyang Liu}, \bibinfo{person}{Yandong Wen},
  \bibinfo{person}{Zhiding Yu}, {and} \bibinfo{person}{Meng Yang}.}
  \bibinfo{year}{2016}\natexlab{}.
\newblock \showarticletitle{Large-Margin Softmax Loss for Convolutional Neural
  Networks}. In \bibinfo{booktitle}{\emph{Proceedings of The 33rd International
  Conference on Machine Learning}} \emph{(\bibinfo{series}{Proceedings of
  Machine Learning Research}, Vol.~\bibinfo{volume}{48})},
  \bibfield{editor}{\bibinfo{person}{Maria~Florina Balcan} {and}
  \bibinfo{person}{Kilian~Q. Weinberger}} (Eds.). \bibinfo{publisher}{PMLR},
  \bibinfo{address}{New York, New York, USA}, \bibinfo{pages}{507--516}.
\newblock
\urldef\tempurl%
\url{https://proceedings.mlr.press/v48/liud16.html}
\showURL{%
\tempurl}


\bibitem[\protect\citeauthoryear{Masnadi-Shirazi, Mahadevan, and
  Vasconcelos}{Masnadi-Shirazi et~al\mbox{.}}{2010}]%
        {masnadi2010design}
\bibfield{author}{\bibinfo{person}{Hamed Masnadi-Shirazi},
  \bibinfo{person}{Vijay Mahadevan}, {and} \bibinfo{person}{Nuno Vasconcelos}.}
  \bibinfo{year}{2010}\natexlab{}.
\newblock \showarticletitle{On the design of robust classifiers for computer
  vision}. In \bibinfo{booktitle}{\emph{2010 IEEE Computer Society Conference
  on Computer Vision and Pattern Recognition}}. IEEE,
  \bibinfo{pages}{779--786}.
\newblock


\bibitem[\protect\citeauthoryear{Masnadi-Shirazi and
  Vasconcelos}{Masnadi-Shirazi and Vasconcelos}{2008}]%
        {masnadi2008design}
\bibfield{author}{\bibinfo{person}{Hamed Masnadi-Shirazi} {and}
  \bibinfo{person}{Nuno Vasconcelos}.} \bibinfo{year}{2008}\natexlab{}.
\newblock \showarticletitle{On the design of loss functions for classification:
  theory, robustness to outliers, and SavageBoost}. In
  \bibinfo{booktitle}{\emph{Proceedings of the 21st International Conference on
  Neural Information Processing Systems}}. \bibinfo{pages}{1049--1056}.
\newblock


\bibitem[\protect\citeauthoryear{Mitchell}{Mitchell}{1997}]%
        {mitchell1997machine}
\bibfield{author}{\bibinfo{person}{Tom Mitchell}.}
  \bibinfo{year}{1997}\natexlab{}.
\newblock \bibinfo{booktitle}{\emph{Machine learning}}.
\newblock \bibinfo{publisher}{McGraw hill Burr Ridge}.
\newblock


\bibitem[\protect\citeauthoryear{Pham, Guan, Zoph, Le, and Dean}{Pham
  et~al\mbox{.}}{2018}]%
        {pham2018efficient}
\bibfield{author}{\bibinfo{person}{Hieu Pham}, \bibinfo{person}{Melody Guan},
  \bibinfo{person}{Barret Zoph}, \bibinfo{person}{Quoc Le}, {and}
  \bibinfo{person}{Jeff Dean}.} \bibinfo{year}{2018}\natexlab{}.
\newblock \showarticletitle{Efficient neural architecture search via parameters
  sharing}. In \bibinfo{booktitle}{\emph{International Conference on Machine
  Learning}}. PMLR, \bibinfo{pages}{4095--4104}.
\newblock


\bibitem[\protect\citeauthoryear{Real, Aggarwal, Huang, and Le}{Real
  et~al\mbox{.}}{2019}]%
        {real2019regularized}
\bibfield{author}{\bibinfo{person}{Esteban Real}, \bibinfo{person}{Alok
  Aggarwal}, \bibinfo{person}{Yanping Huang}, {and} \bibinfo{person}{Quoc~V.
  Le}.} \bibinfo{year}{2019}\natexlab{}.
\newblock \showarticletitle{Regularized Evolution for Image Classifier
  Architecture Search}.
\newblock \bibinfo{journal}{\emph{Proceedings of the AAAI Conference on
  Artificial Intelligence}} \bibinfo{volume}{33}, \bibinfo{number}{01}
  (\bibinfo{date}{Jul.} \bibinfo{year}{2019}), \bibinfo{pages}{4780--4789}.
\newblock
\urldef\tempurl%
\url{https://doi.org/10.1609/aaai.v33i01.33014780}
\showDOI{\tempurl}


\bibitem[\protect\citeauthoryear{Real, Moore, Selle, Saxena, Suematsu, Tan, Le,
  and Kurakin}{Real et~al\mbox{.}}{2017}]%
        {real2017large}
\bibfield{author}{\bibinfo{person}{Esteban Real}, \bibinfo{person}{Sherry
  Moore}, \bibinfo{person}{Andrew Selle}, \bibinfo{person}{Saurabh Saxena},
  \bibinfo{person}{Yutaka~Leon Suematsu}, \bibinfo{person}{Jie Tan},
  \bibinfo{person}{Quoc~V Le}, {and} \bibinfo{person}{Alexey Kurakin}.}
  \bibinfo{year}{2017}\natexlab{}.
\newblock \showarticletitle{Large-scale evolution of image classifiers}. In
  \bibinfo{booktitle}{\emph{International Conference on Machine Learning}}.
  PMLR, \bibinfo{pages}{2902--2911}.
\newblock


\bibitem[\protect\citeauthoryear{Rosasco, De~Vito, Caponnetto, Piana, and
  Verri}{Rosasco et~al\mbox{.}}{2004}]%
        {rosasco2004loss}
\bibfield{author}{\bibinfo{person}{Lorenzo Rosasco}, \bibinfo{person}{Ernesto
  De~Vito}, \bibinfo{person}{Andrea Caponnetto}, \bibinfo{person}{Michele
  Piana}, {and} \bibinfo{person}{Alessandro Verri}.}
  \bibinfo{year}{2004}\natexlab{}.
\newblock \showarticletitle{Are loss functions all the same?}
\newblock \bibinfo{journal}{\emph{Neural computation}} \bibinfo{volume}{16},
  \bibinfo{number}{5} (\bibinfo{year}{2004}), \bibinfo{pages}{1063--1076}.
\newblock


\bibitem[\protect\citeauthoryear{Rubinstein}{Rubinstein}{1999}]%
        {rubinstein1999cross}
\bibfield{author}{\bibinfo{person}{Reuven Rubinstein}.}
  \bibinfo{year}{1999}\natexlab{}.
\newblock \showarticletitle{The cross-entropy method for combinatorial and
  continuous optimization}.
\newblock \bibinfo{journal}{\emph{Methodology and computing in applied
  probability}} \bibinfo{volume}{1}, \bibinfo{number}{2}
  (\bibinfo{year}{1999}), \bibinfo{pages}{127--190}.
\newblock


\bibitem[\protect\citeauthoryear{Shi, Chen, Ma, Mao, Zhang, and Zhang}{Shi
  et~al\mbox{.}}{2020}]%
        {shi2020neural}
\bibfield{author}{\bibinfo{person}{Shaoyun Shi}, \bibinfo{person}{Hanxiong
  Chen}, \bibinfo{person}{Weizhi Ma}, \bibinfo{person}{Jiaxin Mao},
  \bibinfo{person}{Min Zhang}, {and} \bibinfo{person}{Yongfeng Zhang}.}
  \bibinfo{year}{2020}\natexlab{}.
\newblock \showarticletitle{Neural Logic Reasoning}. In
  \bibinfo{booktitle}{\emph{Proceedings of the 29th ACM International
  Conference on Information \& Knowledge Management}}.
  \bibinfo{pages}{1365--1374}.
\newblock


\bibitem[\protect\citeauthoryear{Thomas~Elsken}{Thomas~Elsken}{2018}]%
        {elsken2017simple}
\bibfield{author}{\bibinfo{person}{Frank~Hutter Thomas~Elsken, Jan
  Hendrik~Metzen}.} \bibinfo{year}{2018}\natexlab{}.
\newblock \bibinfo{title}{Simple and efficient architecture search for
  Convolutional Neural Networks}.
\newblock
\newblock
\urldef\tempurl%
\url{https://openreview.net/forum?id=SySaJ0xCZ}
\showURL{%
\tempurl}


\bibitem[\protect\citeauthoryear{Wang, Wang, Chi, Zhang, and Mei}{Wang
  et~al\mbox{.}}{2020}]%
        {wang2020loss}
\bibfield{author}{\bibinfo{person}{Xiaobo Wang}, \bibinfo{person}{Shuo Wang},
  \bibinfo{person}{Cheng Chi}, \bibinfo{person}{Shifeng Zhang}, {and}
  \bibinfo{person}{Tao Mei}.} \bibinfo{year}{2020}\natexlab{}.
\newblock \showarticletitle{Loss function search for face recognition}. In
  \bibinfo{booktitle}{\emph{International Conference on Machine Learning}}.
  PMLR, \bibinfo{pages}{10029--10038}.
\newblock


\bibitem[\protect\citeauthoryear{Williams}{Williams}{1992}]%
        {williams1992simple}
\bibfield{author}{\bibinfo{person}{Ronald~J Williams}.}
  \bibinfo{year}{1992}\natexlab{}.
\newblock \showarticletitle{Simple statistical gradient-following algorithms
  for connectionist reinforcement learning}.
\newblock \bibinfo{journal}{\emph{Machine learning}} \bibinfo{volume}{8},
  \bibinfo{number}{3} (\bibinfo{year}{1992}), \bibinfo{pages}{229--256}.
\newblock


\bibitem[\protect\citeauthoryear{Willmott and Matsuura}{Willmott and
  Matsuura}{2005}]%
        {willmott2005advantages}
\bibfield{author}{\bibinfo{person}{Cort~J Willmott} {and}
  \bibinfo{person}{Kenji Matsuura}.} \bibinfo{year}{2005}\natexlab{}.
\newblock \showarticletitle{Advantages of the mean absolute error (MAE) over
  the root mean square error (RMSE) in assessing average model performance}.
\newblock \bibinfo{journal}{\emph{Climate research}} \bibinfo{volume}{30},
  \bibinfo{number}{1} (\bibinfo{year}{2005}), \bibinfo{pages}{79--82}.
\newblock


\bibitem[\protect\citeauthoryear{Xu, Zhang, Hu, Liang, Salakhutdinov, and
  Xing}{Xu et~al\mbox{.}}{2019}]%
        {xu2018autoloss}
\bibfield{author}{\bibinfo{person}{Haowen Xu}, \bibinfo{person}{Hao Zhang},
  \bibinfo{person}{Zhiting Hu}, \bibinfo{person}{Xiaodan Liang},
  \bibinfo{person}{Ruslan Salakhutdinov}, {and} \bibinfo{person}{Eric Xing}.}
  \bibinfo{year}{2019}\natexlab{}.
\newblock \showarticletitle{AutoLoss: Learning Discrete Schedule for Alternate
  Optimization}. In \bibinfo{booktitle}{\emph{International Conference on
  Learning Representations}}.
\newblock
\urldef\tempurl%
\url{https://openreview.net/forum?id=BJgK6iA5KX}
\showURL{%
\tempurl}


\bibitem[\protect\citeauthoryear{Xu, Ge, Li, Fu, Chen, and Zhang}{Xu
  et~al\mbox{.}}{2021}]%
        {xu2021causal}
\bibfield{author}{\bibinfo{person}{Shuyuan Xu}, \bibinfo{person}{Yingqiang Ge},
  \bibinfo{person}{Yunqi Li}, \bibinfo{person}{Zuohui Fu}, \bibinfo{person}{Xu
  Chen}, {and} \bibinfo{person}{Yongfeng Zhang}.}
  \bibinfo{year}{2021}\natexlab{}.
\newblock \showarticletitle{Causal collaborative filtering}.
\newblock \bibinfo{journal}{\emph{arXiv preprint arXiv:2102.01868}}
  (\bibinfo{year}{2021}).
\newblock


\bibitem[\protect\citeauthoryear{Yao, Wang, Chen, Dai, Li, Tu, Yang, and
  Yu}{Yao et~al\mbox{.}}{2018}]%
        {yao2018taking}
\bibfield{author}{\bibinfo{person}{Quanming Yao}, \bibinfo{person}{Mengshuo
  Wang}, \bibinfo{person}{Yuqiang Chen}, \bibinfo{person}{Wenyuan Dai},
  \bibinfo{person}{Yu-Feng Li}, \bibinfo{person}{Wei-Wei Tu},
  \bibinfo{person}{Qiang Yang}, {and} \bibinfo{person}{Yang Yu}.}
  \bibinfo{year}{2018}\natexlab{}.
\newblock \showarticletitle{Taking human out of learning applications: A survey
  on automated machine learning}.
\newblock \bibinfo{journal}{\emph{arXiv preprint arXiv:1810.13306}}
  (\bibinfo{year}{2018}).
\newblock


\bibitem[\protect\citeauthoryear{Zhao, Liu, Fan, Liu, Tang, and Wang}{Zhao
  et~al\mbox{.}}{2021}]%
        {zhao2021autoloss}
\bibfield{author}{\bibinfo{person}{Xiangyu Zhao}, \bibinfo{person}{Haochen
  Liu}, \bibinfo{person}{Wenqi Fan}, \bibinfo{person}{Hui Liu},
  \bibinfo{person}{Jiliang Tang}, {and} \bibinfo{person}{Chong Wang}.}
  \bibinfo{year}{2021}\natexlab{}.
\newblock \showarticletitle{AutoLoss: Automated Loss Function Search in
  Recommendations}. In \bibinfo{booktitle}{\emph{{KDD} '21: The 27th {ACM}
  {SIGKDD} Conference on Knowledge Discovery and Data Mining, Virtual Event,
  Singapore, August 14-18, 2021}}, \bibfield{editor}{\bibinfo{person}{Feida
  Zhu}, \bibinfo{person}{Beng~Chin Ooi}, {and} \bibinfo{person}{Chunyan Miao}}
  (Eds.). \bibinfo{publisher}{{ACM}}, \bibinfo{pages}{3959--3967}.
\newblock
\urldef\tempurl%
\url{https://doi.org/10.1145/3447548.3467208}
\showDOI{\tempurl}


\bibitem[\protect\citeauthoryear{Zoph and Le}{Zoph and Le}{2017}]%
        {zoph2016neural}
\bibfield{author}{\bibinfo{person}{Barret Zoph} {and} \bibinfo{person}{Quoc~V.
  Le}.} \bibinfo{year}{2017}\natexlab{}.
\newblock \showarticletitle{Neural Architecture Search with Reinforcement
  Learning}. In \bibinfo{booktitle}{\emph{5th International Conference on
  Learning Representations, {ICLR} 2017, Toulon, France, April 24-26, 2017,
  Conference Track Proceedings}}. \bibinfo{publisher}{OpenReview.net}.
\newblock
\urldef\tempurl%
\url{https://openreview.net/forum?id=r1Ue8Hcxg}
\showURL{%
\tempurl}


\bibitem[\protect\citeauthoryear{Zoph, Vasudevan, Shlens, and Le}{Zoph
  et~al\mbox{.}}{2018}]%
        {zoph2018learning}
\bibfield{author}{\bibinfo{person}{Barret Zoph}, \bibinfo{person}{Vijay
  Vasudevan}, \bibinfo{person}{Jonathon Shlens}, {and} \bibinfo{person}{Quoc~V
  Le}.} \bibinfo{year}{2018}\natexlab{}.
\newblock \showarticletitle{Learning transferable architectures for scalable
  image recognition}. In \bibinfo{booktitle}{\emph{Proceedings of the IEEE
  conference on computer vision and pattern recognition}}.
  \bibinfo{pages}{8697--8710}.
\newblock


\end{thebibliography}

\end{document}